\documentclass[12pt]{article}
\usepackage{comment}
\usepackage{amsmath}
\usepackage{amssymb}
\usepackage{graphicx}
\usepackage{here}
\usepackage{subcaption}
\usepackage{url}
\usepackage[sort&compress, numbers, merge]{natbib}
\usepackage{braket}
\usepackage{physics}
\usepackage[compat=1.1.0]{tikz-feynhand}
\usepackage{slashed}
\usepackage{mathtools}
\usepackage{tikz}
\usetikzlibrary{decorations.markings}
\usepackage{cancel}

\usepackage{todonotes}

\numberwithin{equation}{section}

\usepackage[colorlinks=true, linkcolor=blue, citecolor=blue,
urlcolor=black]{hyperref}

\begin{document}
\baselineskip 0.6cm
\def\simgt{\mathrel{\lower2.5pt\vbox{\lineskip=0pt\baselineskip=0pt
           \hbox{$>$}\hbox{$\sim$}}}}
\def\simlt{\mathrel{\lower2.5pt\vbox{\lineskip=0pt\baselineskip=0pt
           \hbox{$<$}\hbox{$\sim$}}}}
\def\simprop{\mathrel{\lower3.0pt\vbox{\lineskip=1.0pt\baselineskip=0pt
             \hbox{$\propto$}\hbox{$\sim$}}}}

\def\SU{\mathrm{SU}}
\def\SM{\mathrm{SM}}
\def\FN{\mathrm{FN}}
\def\ubar{\overline{u}}
\def\dbar{\overline{d}}
\def\ebar{\overline{e}}
\def\Lbar{\overline{L}}
\def\Ebar{\overline{E}}
\def\Nbar{\overline{N}}
\def\FN{\mathrm{FN}}
\def\CKM{\mathrm{CKM}}
\def\PMNS{\mathrm{PMNS}}
\def\calL{\mathcal{L}}
\def\fivebar{\overline{\mathbf{5}}}
\def\ten{\mathbf{10}}
\def\calN{\mathcal{N}}
\def\Re{\mathrm{Re}}
\def\Im{\mathrm{Im}}
\def\MSbar{\overline{\mathrm{MS}}}
\def\UFN{\mathrm{U(1)_{FN}}}

\begin{titlepage}

\begin{flushright}
IPMU24-0039
\end{flushright}

\vskip 1.1cm

\begin{center}

{\Large \bf 
Nucleon Decay as a Probe of Flavor Symmetry:\\
The Case of Fake Unification 
}

\vskip 1.2cm

\vskip 1.2cm
Masahiro Ibe$^{a,b}$, 
Satoshi Shirai$^{b}$ and 
Keiichi Watanabe$^{a}$ 
\vskip 0.5cm

{\it

$^a$ {ICRR, The University of Tokyo, Kashiwa, Chiba 277-8582, Japan}

$^b$ {Kavli Institute for the Physics and Mathematics of the Universe
(WPI), \\The University of Tokyo Institutes for Advanced Study, \\ The
University of Tokyo, Kashiwa 277-8583, Japan}
}

\vskip 1.0cm

\abstract{
This paper explores nucleon decay within the framework of a ``fake Grand Unified Theory (GUT)" combined with the Froggatt-Nielsen (FN) mechanism. 
In this fake GUT framework, 
quarks and leptons may have distinct high-energy origins but fit into complete $\SU(5)$ multiplets at low energies without requiring force unification, 
setting it apart from conventional GUTs. 
By introducing flavor symmetry through the FN mechanism, 
the model addresses the flavor puzzle of quark and lepton mass hierarchies and mixing patterns. 
Our analysis demonstrates that nucleon decay rates and branching fractions in the fake GUT are sensitive to flavor symmetry, 
providing a means to distinguish it from conventional GUT predictions. 
These findings underscore the importance of nucleon decay searches in probing both baryon number violation and the underlying flavor structure.
}

\end{center}
\end{titlepage}

\section{Introduction} 

The Standard Model (SM) is remarkably successful due to its 
precise predictions across fundamental interactions, many of which have been experimentally confirmed.
However, 
there remains the question of why quarks and leptons possess specific gauge representations.
Interestingly, 
the matter fields of the SM have gauge representations that can be neatly embedded into the $\fivebar$ and $\ten$ representations of SU(5).
Grand Unified Theory (GUT) is a framework which can explain this apparent matter unification 
by embedding the SM 
gauge groups into the SU(5) gauge group.
Consequently,
the study of the GUT has been a long-standing pursuit\,\cite{Georgi:1974sy,Georgi:1974yf,Buras:1977yy} (see Ref.\,\cite{ParticleDataGroup:2024cfk} for reviews).
The GUT predicts nucleon decay,
which is being extensively searched for across a variety of current and future experiments. (see e.g., Refs.\,\cite{Super-Kamiokande:2020wjk,Hyper-Kamiokande:2018ofw,JUNO:2015zny,DUNE:2020fgq}).

To explain the apparent matter unification, however, 
the force unification of the SM is not mandatory.
In fact, the framework called ``fake GUT" has been proposed as an alternative to explain the unification of SM matter into the SU(5) multiplets without force unification\,\cite{Ibe:2019ifm,Ibe:2022ock}
(see also Refs.\,\cite{Hall:2018let,Hall:2019qwx} for related discussions within the SO(10) GUT model). 
In this framework, quarks and leptons may have distinct origins at high energies although they automatically fit into complete $\mathrm{SU}(5)$ multiplets at low energies.
This approach helps address challenges faced by conventional GUT models,
such as too rapid proton decay, 
failure to achieve successful gauge coupling unification, 
or difficulty in unifying quark and lepton Yukawa couplings.
A notable aspect of the fake GUT is its ability to yield predictions for nucleon decay that vary significantly based on the flavor structure of the matter fields, as discussed in Ref.\,\cite{Ibe:2022ock}.

The flavor puzzle is another question that cannot be explained within the SM.
This puzzle encompasses unresolved questions regarding the mass hierarchy of quarks and leptons, the specific structure of the CKM matrix, and flavor mixing of neutrinos. 
It remains quite puzzling that, although there is no distinction between the three generations of fermions from the perspective of gauge interactions, their Yukawa interactions exhibit significant differences.
To address these  issues,
various flavor models have been 
proposed in addition to the GUT framework.

To explain the above flavor structures, 
various models and ideas have been proposed (see e.g., Refs.\,\cite{Fritzsch:1999ee,Altarelli:2010gt} for reviews).
Among these,
the Froggatt-Nielsen (FN) mechanism provides a simple approach which can solve the flavor puzzle by utilizing the U(1) symmetry and its subsequent breaking\,\cite{Froggatt:1978nt}. 
With appropriate U(1) charge assignments of the SM fermions,
hierarchical effective Yukawa interactions can be achieved 
from featureless Yukawa interactions where the fundamental dimensionless couplings are of order unity.
This framework also allows for the reproduction of the observed mixing patterns across generations.

In this paper, 
we analyze the nucleon decay in the fake GUT with FN mechanism.
As we will demonstrate, 
the nucleon decay rate and its branching fractions are sensitive to flavor symmetry in the fake GUT scenario. 
This contrasts sharply with 
nucleon decay mediated by GUT gauge bosons in conventional GUT models, 
where 
decay characteristics remain largely unaffected by flavor symmetry.
This feature highlights
the experimental potential to differentiate between fake GUT and conventional GUT models in experiments such as Super-Kamiokande (SK), Hyper-Kamiokande (HK),
JUNO, and DUNE experiments\,\cite{Super-Kamiokande:2020wjk,Hyper-Kamiokande:2018ofw,JUNO:2015zny,DUNE:2020fgq}. 
Additionally, 
these nucleon decay experiments offer a unique opportunity to probe aspects of flavor symmetry.

The organization of this paper is as follows.
We first summarize the concrete fake GUT model based on $\SU(5)\times \mathrm{U}(2)_H$ in Ref.\,\cite{Ibe:2022ock} in Sec.\,\ref{sec:u2_fakegut}.
In Sec.\,\ref{sec:FNmechanism}, 
we introduce the FN mechanism, 
and we examine several FN charge assignments that successfully 
reproduce the flavor structure of the SM in the fake GUT model.
In Sec.\,\ref{sec:fake_flavor}, 
we investigate the nucleon decay in the $\SU(5)\times \mathrm{U}(2)_H$ model with the FN mechanism.
In the final section, 
we present our conclusions.

\section{Fake GUT Model Based on \texorpdfstring{$\boldsymbol{\SU(5) \times \mathrm{U}(2)_H}$}{}}
\label{sec:u2_fakegut}

In this section,
we summarize a minimal fake GUT model based on $G = \SU(5) \times \mathrm{U}(2)_{H} \simeq \SU(5) \times \SU(2)_H \times \mathrm{U}(1)_{H}$ in Ref.\,\cite{Ibe:2022ock}.
Specifically, 
we provide a brief overview of the origin of SM fermions and Yukawa interactions,
as well as key features of proton decay.

\subsection{Origin of SM fermions}
\label{sec:origin_fermions}

In this model, 
there are three generations of chiral fermions $\fivebar \oplus \ten $ of $\SU(5)$, 
and 
three pairs of vector-like fermions charged under U(2)$_{H}$ as follows:%
\footnote{In this paper, we use Weyl fermion notation throughout.}
\begin{gather}
\label{eq:U2 lepton doublet}
\fivebar:
(\fivebar,\mathbf{1})_{0}\,\times\, 3\ , \quad
\ten:(\ten,\mathbf{1})_{0}\, \times \, 3\ , \\
[\, L_{H}:(\mathbf{1},\mathbf{2})_{-1/2}\ , \quad 
\Lbar_{H}:(\mathbf{1},\mathbf{2})_{+1/2}\, ]\,\times\, 3\ , \\
\label{eq:U2 lepton singlet}
[\, E_{H}:(\mathbf{1},\mathbf{1})_{-1}\ , \quad \Ebar_{H}:(\mathbf{1},\mathbf{1})_{+1} \,]\,\times\, 3\ .
\end{gather}
Here, 
group representations are denoted by $(\SU(5), \,\SU(2)_{H})_{\mathrm{U}(1)_{H}}$.
As discussed later,
while the SM quark sector originates fully  from $\fivebar \oplus \ten$, 
the SM lepton sector is derived from a linear combination of $\fivebar\,(\ten)$ and $L_H\, (\Ebar_H)$.

The spontaneous breaking
of $\SU(5)\times \mathrm{U}(2)_H$ into $G_{\mathrm{SM}} = \SU(3)_c \times \SU(2)_L \times \mathrm{U}(1)_Y$
is achieved by a vacuum expectation value (VEV) of a complex scalar field $\phi_{2}$,
which is in the bi-fundamental representation, 
$(\mathbf{5}, \mathbf{2})_{-1/2}$.
Explicitly, 
the VEV
of $\phi_{2}$
is given by
\begin{align}
\label{eq: Phi vev} 
\ev{\phi_{2}} = 
 \left( \begin{array}{ccccc}
      0 & 0 & 0 & v_{2} & 0 \\
      0 & 0 & 0 & 0 & v_{2}
    \end{array} \right)\ ,
\end{align}
which breaks $\SU(5)\times \mathrm{U}(2)_H$ down to $G_{\mathrm{SM}}$.
Here, 
$v_2$ is a scale associated with the fake GUT,  much larger than the electroweak scale.
In this case, 
$\SU(3)_{c}$ remains as an unbroken subgroup of $\SU(5)$,
while $\SU(2)_{L}$ and $\mathrm{U}(1)_{Y}$ appear as diagonal subgroups of $\SU(5)$ and $\mathrm{U}(2)_H$.
Additionally,
$\fivebar$ and $\ten$ decompose as follows:
\begin{align}
\fivebar \rightarrow \dbar_{\SM} \oplus L_{\fivebar}\ , \quad
\ten \rightarrow Q_{\SM} \oplus \ubar_{\SM} \oplus \Ebar_{\ten}\ .
\end{align}
In this model,
SM leptons come from massless linear combinations of $L_{\fivebar} \,(\Ebar_{\ten})$ and $L_H\, (\Ebar_H)$.
To illustrate this point explicitly,
let us consider the following interactions 
between the fermions and $\phi_2$,
\begin{align}
\label{eq:Abelian_mass}
\calL 
&= m_{L,ij}\, \Lbar_{Hi}\, L_{Hj} 
+ \lambda_{L,ij}\, \Lbar_{Hi}\, \phi_2\, \fivebar_{j} 
+ m_{E,ij}\, E_{Hi}\, \Ebar_{Hj} 
+ \frac{\lambda_{E,ij}}{\Lambda_{\mathrm{cut}}}\, E_{Hi}\,
\phi_2^{\dagger}\, \phi_2^{\dagger}\, \ten_j + h.c.\ .
\end{align}
Here, $\lambda_{L,E}$ denote coupling constants, 
and
$\Lambda_{\mathrm{cut}}$ represents a cutoff scale
greater than or comparable to the fake GUT scale\,\cite{Ibe:2022ock}.
The flavor indices are denoted by $i$ and $j$.
The higher-dimensional operator arises by integrating out fields heavier than the fake GUT scale
(see Ref.\,\cite{Ibe:2022ock} for details).

After the fake GUT symmetry breaking,
the mass terms for the leptonic components  take the form:
\begin{align}
\label{eq:leptonic_mass}
\mathcal{L}_{\mathrm{mass}}
  = \Lbar_{Hi}
  \,\mathcal{M}_{L,ij}
   \left(
    \begin{array}{c}
      L_{\fivebar} \\
      L_{H}
    \end{array}
  \right)_j + 
  E_{Hi} \,\mathcal{M}_{E,ij}
  \left(
    \begin{array}{c}
      \Ebar_{\ten} \\
      \Ebar_{H}
    \end{array}
  \right)_{j} +
  h.c.\ ,
\end{align}
where $\mathcal{M}_{L,E}$ are the $3 \times 6$ mass matrices, 
given by:
\begin{align}
\label{eq:leptonic_mass_matrix}
     \mathcal{M}_{L} = 
        \left(
    \begin{array}{cc}
      \lambda_{L}\, v_2 & m_{L}
    \end{array}
  \right)\ , \quad 
     \mathcal{M}_{E} = 
         \left(
     \begin{array}{cc}
    \displaystyle{\frac{\lambda_{E}\, v_2^2}{\Lambda_{\mathrm{cut}}}} & m_{E}
     \end{array}
    \right)\ .
\end{align}
Three leptons remain massless due to the rank conditions of these $3 \times 6$ matrices.
As a result, 
the three generations of Weyl fermions of the SM are realized,
although the SM leptons are not fully contained within 
$\fivebar\oplus\ten$ unlike conventional SU(5) GUT models.
On the other hand, 
all SM quarks are fully incorporated within $\fivebar\oplus\ten$.

To explicitly show the emergence of SM leptons at low energy,
let us ignore flavor mixing and focus on just one generation for simplicity.
In this case, 
the mass eigenstates of the leptonic components are given by 
\begin{align}
\label{eq:L_mixing}
  &\left(
    \begin{array}{c}
      L_M \\
      \ell_{\SM}
    \end{array}
  \right) 
  = \left(
    \begin{array}{cc}
      \mathrm{cos}\, \theta_L & \mathrm{sin}\, \theta_L \\
      -\mathrm{sin}\, \theta_L & \mathrm{cos}\, \theta_L
    \end{array}
  \right) 
  \left(
    \begin{array}{c}
      L_{\fivebar} \\
      L_{H}
    \end{array}
  \right)\ , \\
  \label{eq:E_mixing}
  &\left(
    \begin{array}{c}
      \Ebar_{M} \\
      \ebar_{\SM}
    \end{array}
  \right) 
  = \left(
    \begin{array}{cc}
      \mathrm{cos}\, \theta_{E} & \mathrm{sin}\, \theta_{E} \\
      -\mathrm{sin}\, \theta_{E} & \mathrm{cos}\, \theta_{E}
    \end{array}
  \right) 
  \left(
    \begin{array}{c}
      \Ebar_{\ten} \\
      \Ebar_{H}
    \end{array}
  \right)\ .
  \end{align}
Here, $\ell_{\SM}$ and $\ebar_{\SM}$ represent the massless eigenstates of the doublet and the singlet, respectively.
Additionally,
the leptonic mixing angles, $\theta_{L,E}$, are defined by
\begin{align}
\label{eq:mixing_angle}
\mathrm{tan}\, \theta_{L} 
= \qty|\frac{m_{L}}{\lambda_{L}\, v_{2}}|\ , \quad
\mathrm{tan}\, \theta_{E} 
= \qty|\frac{ m_{E} \, \Lambda_{\mathrm{cut}}}{\lambda_{E}\, v_{2}^{2}}|\ .
\end{align}
In this way, 
we obtain the SM leptons and quarks. 
Notably, 
the resultant fermions automatically fit into  $\fivebar \oplus \ten$ multiplets, despite not being embedded in a true GUT multiplet. 
We refer to this mechanism as the fake GUT,
arising from the chiral structure of the SU(5) sector.

Note that in the limit where $m_{L,E}\to 0$, a global symmetry is enhanced, under which $L_H$ and $\Ebar_H$ carry distinct charges.
Thus, it is technically natural 
to assume that $m_{L,E}$ are small
in the sense of 't~Hooft's naturalness criterion\,\cite{tHooft:1980xss}.
In the small $m_{L,E}$ limit,
SM leptons are fully contained in 
$L_H$ and $\Ebar_H$,
while the heavy leptons, 
$L_M$ and $\Ebar_M$, remain massive.
In this case, although the quarks and leptons in the SM originate from 
completely separate multiplets in the fake GUT, 
they appear to form 
$\fivebar \oplus \ten$ multiplets at low energies.
For future reference,
we refer to this global symmetry as $\mathrm{U}(1)_{LH}$ symmetry under which $L_H$ and $\Ebar_H$ have charges, $+1$ and $-1$, respectively.

\subsection{Origin of SM Higgs and Yukawa interactions}
\label{sec:origin_higgs}

As discussed in the previous section, 
since the SM quarks and the SM leptons can have different UV origins,
the SM Yukawa interactions are composed of various contributions.
In this section, we show a concrete example of the origin of the SM Yukawa 
interactions.
For that purpose, we introduce a scalar field, 
$H_{2}$, which is in the $(\mathbf{1}, \mathbf{2})_{1/2}$ representation.
We also introduce $H_{5}$ in the $(\mathbf{5}, \mathbf{1})_{0}$ representation,
which decomposes as the triplet and doublet,
\begin{align}
H_5 
= H_C \oplus H_1\ .
\end{align}

The two Higgs doublets, 
$H_1$ and $H_2$,
are mixed by the following interaction,
\begin{align}
\label{eq:Higgs_mixing}
\calL_{\mathrm{mix}} 
= 
\mu_{\mathrm{mix}}\, H_2\, \phi_2\, H_5^{\dagger} + h.c.\ ,
\end{align}
where $\mu_{\mathrm{mix}}$ is a mass parameter of $\order{v_2}$.
After $\SU(5)\times \mathrm{U}(2)_H$ symmetry breaking,
the Higgs doublets obtain the following effective mass terms,
\begin{align}
\calL
= 
- m_1^2\, |H_1|^2 
- m_2^2\, |H_2|^2 
+ (\mu_{\mathrm{mix}}\, v_2\, H_2\, H_1^{\dagger} + h.c.)\ .
\end{align}
Here, $m_1^2$ and $m_2^2$
are mass parameters of $\order{v_{2}^2}$.
The SM Higgs, $h_{\SM}$, is given by an almost massless linear combination of $H_1$ and $H_2$,
\begin{align}
\label{eq:higgs_mixing}
h_{\SM}
=
\sin\theta_h\, H_1
+ \cos\theta_h\, H_2\ .
\end{align}
To achieve the mass term of 
$h_{\SM}$ in $\order{100}$\,GeV, 
we require fine-tuning
similar to conventional GUT models.

By using $H_5$ and $H_2$, 
the UV origins of the SM Yukawa interactions are given by,
\begin{align}
\label{eq:quark Yukawa}
\calL_{YQ} 
&= 
\frac{1}{4}\, y^{(10)}_{ij}\, \ten_i\, \ten_j\, H_5
-\sqrt{2}\, y^{(5)}_{ij}\, 
\ten_i\, \fivebar_j\, 
H_5^{\dagger} + h.c.\ , \\
\label{eq:lepton Yukawa}
\calL_{YL} 
&= -y^{(LE)}_{ij} \, L_{Hi}\, \Ebar_{Hj}\, H_2^{\dagger}
+ h.c.\ ,
\end{align}
where $i$ and $j$
represent flavor indices.
The parameters
$y^{(10)}_{ij}$, $y^{(5)}_{ij}$ and $y^{(LE)}_{ij}$
are dimensionless couplings.
The SM Yukawa couplings are obtained by 
substituting $H_5 \to  \sin\theta_h\, h_{\SM}$, $H_2 \to \cos\theta_h\, h_{\SM}$ after diagonalizing the mass matrices in Eq.\,\eqref{eq:leptonic_mass_matrix}.
For simplicity, we will omit the $\SM$ subscript when the SM fields are clear from context.

As we will discuss in Sec.\,\ref{sec:Nucleondecay_review},
the lepton components in $\fivebar\oplus \ten$ must be highly suppressed 
to avoid nucleon decay constraints,
i.e., $|\theta_{L,E}| \ll 1$.
In this case, 
the SM Yukawa interactions are given as, 
\begin{align}
\label{eq:SM_yukawa}   
\calL_Y 
= 
- y^{(u)}_{ij}\, Q_{i}\, \ubar_{j}\, h
- y^{(d)}_{ij}\, Q_{i}\, \dbar_{j}\, h^{\dagger}
- y^{(e)}_{ij}\, 
\ell_{i}\, \ebar_{j}\, h^{\dagger}
+ h.c.\ ,
\end{align}
where the SM Yukawa couplings are written by
\begin{align}
\label{eq:Yukawa_Origins}
\begin{split}
y^{(u)}_{ij} 
=& \sin\theta_h\, y^{(10)}_{ij}\ , \\
y^{(d)}_{ij}
=& \sin\theta_h\, y^{(5)}_{ij}\ , \\
y^{(e)}_{ij}
=& \cos\theta_h\, y^{(LE)}_{ij}
+ \order{\theta_{L}\theta_{E}} 
\sin\theta_h\, y^{(5)}_{ji}\ .
\end{split}
\end{align}
Here, 
$Q =(u,d)$, $\ubar$, $\dbar$ 
are the doublet, 
the anti-up-type, 
and the anti-down-type quarks, while $\ell=(\nu,e)$, 
$\ebar$ are the doublet and the anti-electron-type leptons
with $i,j$ as flavor indices.
Since we assume $|\theta_{L,E}| \ll 10^{-4}$ to avoid too rapid proton decay,
the $y^{(5)}$ contribution to $y^{(e)}_{ij}$ is negligible.

Finally, 
we describe the origin of the neutrino masses.
In this model, 
we introduce three right-handed neutrinos, $\Nbar_a$, $a=1,2,3$.
Following the seesaw mechanism\,\cite{Minkowski:1977sc,Yanagida:1979as,*Yanagida:1979gs,Gell-Mann:1979vob,Glashow:1979nm,Mohapatra:1979ia},
the origins of the Yukawa interactions to the right-handed neutrinos are given by,%
\footnote{In general,
an another Yukawa interaction, $ y^{(5N)}_{ia}\, \fivebar_i\, \Nbar_a\, H_5 $, exists.
In this paper,
we do not consider this term.}
\begin{align}
\label{eq:all_nbar_int}
\calL_{\nu} 
= 
- y^{(LN)}_{ia}\, L_{Hi}\, \Nbar_{a}\, H_2
- \frac{1}{2}\, M_{R}\, y_{ab}^{(R)}\, \Nbar_{a}\, \Nbar_{b} 
+ h.c.
\end{align}
Here, $M_R$ represents 
the mass scale of the right-handed neutrinos,
and 
the parameters 
$y^{(5N)}_{ia}$, $y^{(LN)}_{ia}$ and $y_{ab}^{(R)}$ are dimensionless coupling constants.
Since we assume
$|\theta_{L,E}| \ll 1$, 
the Yukawa interaction of the right-handed neutrinos is reduced to,
\begin{align}
\label{eq:seesaw_LY}
\calL_{\nu} 
= 
- y^{(D)}_{ia}\, \ell_i\, \Nbar_a\, h
- \frac{1}{2}\, M_R\, y_{ab}^{(R)}\, \Nbar_a\, \Nbar_b + h.c.\ ,
\end{align}
where the Yukawa couplings of the right-handed neutrino are expressed by
\begin{align}
y^{(D)}_{ij} 
= \cos\theta_h\, y^{(LN)}_{ij}\ .
\end{align}

Below the right-handed neutrino mass scale, 
the Yukawa interaction in Eq.\,\eqref{eq:seesaw_LY}
induces effective dimension-five operators
via the seesaw mechanism\,\cite{Minkowski:1977sc,Yanagida:1979as,*Yanagida:1979gs,Gell-Mann:1979vob,Glashow:1979nm,Mohapatra:1979ia},
\begin{gather}
\label{eq:seesaw_dim5}
\calL_{\nu} 
= 
- y^{(S)}_{ij}\,
\frac{(\ell_i\, h) (\ell_j\, h)}{2M_R} 
+ h.c.\ , \quad
y^{(S)}_{ij} 
= 
-(y^{(D)} (y^{(R)})^{-1}\, 
(y^{(D)})^T)_{ij} \ .
\end{gather}
With the VEV of the Higgs doublet, 
$\langle h \rangle = (0,v_\mathrm{EW})^T$,
the neutrinos $\nu_i$ obtain Majorana mass terms, 
\begin{align}
\calL_{\mathrm{mass}}
=    
- \frac{1}{2}\, m_{ij}^{(\nu)}\, \nu_i\, \nu_j + h.c.\ , \quad
m^{(\nu)}_{ij} 
= y^{(S)}_{ij} \frac{v^{2}_\mathrm{EW}}{M_R} \ .
\end{align}
Here, $v_{\mathrm{EW}} \simeq 174\,\mathrm{GeV}$, represents the electroweak scale.

Before concluding this section,
we comment on $m_{L,E}\to 0$ limit
in the presence of the neutrino interactions in Eq.\,\eqref{eq:all_nbar_int}.
In this limit,
the SM charged leptons 
reside entirely in $L_H$'s, 
and hence,
the presence of neutrino mass requires $y^{(LN)}\neq 0$.
For the seesaw mechanism to work, 
we assume 
$M_R\neq 0$, 
which explicitly breaks the U$(1)_{LH}$ 
symmetry down to $Z_2$ symmetry.
Despite the breaking of U$(1)_{LH}$ symmetry,
proton stability is maintained
for $m_{L,E} = 0$ 
due to the remaining $Z_2$ symmetry.

\subsection{Nucleon decay}
\label{sec:Nucleondecay_review}

\begin{figure}[h!]

\centering
\begin{tikzpicture}[scale=1.5]

\draw[thick] (15.3,-1.2) -- (16.5,0)
node[pos=0.0, above left] {\Large $q$};

\draw[thick] (15.3,1.2) -- (16.5,0)
node[pos=0.0, below left] {\Large $q$};
\filldraw[black] (16.5,0) circle (2pt) node[below right] {};
\draw[thick,decorate,decoration=snake] (16.5,0) -- (18.5,0)
node[midway,above] {\Large $X_{\SU(5)}$};

\filldraw[black] (18.5,0) circle (2pt) node[below right] {};

\draw[thick] (18.5,0) -- (19.7,-1.2)
node[pos=1.0, above right] {\Large $l_{\SM}$};

\draw[thick] (18.5,0) -- (19.7,1.2)
node[pos=1.0, below right] {\Large $q$};

\node at (18.89,-0.75)[rotate=45,text=red,anchor=north west] {\Large $\times$};
\node at (18.5,-0.50){\Large $l_{\SU(5)} $};
\end{tikzpicture}
\caption{Baryon and lepton number-violating interaction through the exchange of a heavy SU(5) gauge boson, $X$.
The red cross mark represents the mixing of $L_{\fivebar}$ ($\Ebar_{\ten}$) with $\ell_{\SM}$ ($\ebar_{\SM}$).}
\label{fig:tikz_dim6}
\end{figure}
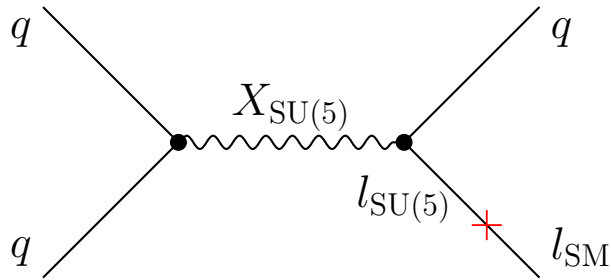

Nucleons can decay for $m_{L,E} \neq 0$
through the exchange of heavy $\SU(5)$ gauge bosons, 
similar to the conventional GUT models.%
\footnote{In Ref.\,\cite{Ibe:2022ock},
we found that there is an upper limit on the heavy SU(5) gauge boson mass, $M_X$, of $M_X^{\mathrm{upper}} \simeq 10^{14.4}\, \mathrm{GeV}$.
This upper limit is derived by requiring 
a consistent matching of the gauge couplings between the fake GUT and the SM at the fake GUT scale,
$\mu_R = M_X$.}
However, 
as discussed in Sec.\,\ref{sec:origin_fermions},
since the origins of the SM quarks and the SM leptons are different,
the interaction related to nucleon decay is suppressed by the leptonic mixing angles, 
$|\mathrm{sin} \, \theta_{L,E}|^2$, 
as shown in Fig.\,\ref{fig:tikz_dim6}.
Let us evaluate the proton lifetime in the simple case where only the first generation leptons have mixing angle $\theta_{L,E}$ in Eqs.\,\eqref{eq:L_mixing} and \eqref{eq:E_mixing}.
In this case, 
the partial proton lifetime for the $ p \rightarrow \pi^{0} e^{+}$ mode is given by
\begin{align}
\label{eq:proton_lifetime_in_U2}
\tau(p \rightarrow \pi^{0} e^{+}) \simeq 
 \frac{5\times 10^{26} \,\mathrm{years}}{ \sin^2 \theta_E + 0.2  \sin^2 \theta_L}\left( \frac{M_{X}/g_{5}}{10^{14}\, \mathrm{GeV}} \right)^{4}\ .
\end{align}
Here, 
$M_X$ and $g_5$ denote the mass of the heavy SU(5) gauge boson and the gauge coupling of SU(5).
This lifetime is consistent with the current experimental limit, 
$\tau(p \rightarrow \pi^{0} e^{+}) > 2.4\times10^{34}$\,years\,\cite{Super-Kamiokande:2020wjk},for small mixing angles,
$|\mathrm{sin}\, \theta_{E,L}| \lesssim 10^{-4}$ even for $M_X = 10^{14}$\,GeV.
Therefore, 
the result in Eq.\,\eqref{eq:proton_lifetime_in_U2}
indicates 
that small mixing angles for both $L_{\fivebar}$ and $\Ebar_{\ten}$ with SM leptons are necessary to satisfy the experimental constraints.%
\footnote{Nucleons can also decay through the exchange of the colored Higgs bosons.
However, 
this process has been shown to be suppressed\,\cite{Ibe:2022ock}.}

In general, 
the lepton mass matrices in Eq.\,\eqref{eq:leptonic_mass_matrix} are flavor-dependent.
In this case,
the flavor composition of the leptonic components $L_{\fivebar, i}$ and $\Ebar_{\ten, i}$ in the $\SU(5)$ sector does not necessarily coincide with the generations of the SM leptons in their mass eigenstates.
Consequently,
the predictions of the nucleon decay rates 
and the branching fractions are different from those in the conventional GUT (see Ref.\,\cite{Ibe:2022ock} for details).
In Sec.\,\ref{sec:fake_flavor}, 
we will discuss predictions for nucleon decay rates and branching fractions within the FN mechanism.

\section{Froggatt-Nielsen Mechanism}
\label{sec:FNmechanism}

\subsection{Explanation of Froggatt-Nielsen mechanism}
\label{sec:summaryFNmechanism}

The FN mechanism is one of the most successful models for explaining the hierarchy of the charged fermion masses and the quark mixing angles\,\cite{Froggatt:1978nt}. 
In this mechanism, a new U(1)  flavor symmetry, $\UFN$,
and a new complex scalar field $\Phi$ are introduced. 
Under the $\UFN$ symmetry, 
quarks and leptons of different generations have distinct $\UFN$ charges, and $\Phi$ has an FN charge of $-1$. 
The $\UFN$ is spontaneously broken when $\Phi$ acquires a VEV, $\langle\Phi\rangle$.%
\footnote{To 
avoid a Goldstone boson associated with spontaneous $\UFN$ breaking, 
we assume the presence of explicit $\UFN$ breaking of a magnitude that does not spoil the FN mechanism.
Alternatively, we may consider 
a discrete $Z_N$ symmetry
with a large order $N$ instead of $\UFN$.}

Let us first consider the SM Yukawa interactions.
Due to the $\UFN$ symmetry,
the Yukawa interactions are modified as
\begin{align}
\label{eq:LY}
\calL_Y 
= 
\kappa_{ij} \left(\frac{\Phi^{(\dagger)}}{M_{*}}\right)^{|f_{\psi,i} + f_{\chi,j}|}\, 
\psi_{i}\, \chi_{j}\, h_{\SM}^{(\dagger)} 
+ h.c. 
\end{align}
Here, $M_{*}$ denotes a cutoff scale, and $\kappa$ represents $3 \times 3$ complex matrices of $\order{1}$. 
The fields $\psi_i$ and $\chi_i$ collectively denote $Q_i$, $L_i$, and $\ubar_i$, $\dbar_i$, $\ebar_i$ (see Eq.\,\eqref{eq:SM_yukawa}). 
The parameters $f_{\psi,i}$ and $f_{\chi,i}$ denote the FN charges of the quarks and leptons, 
and we assume that the Higgs boson does not carry any FN charge.
Due to the locality of the effective field theory,
$\calL_Y$ includes only positive integer powers of $\Phi$ when $f_{\psi,i} + f_{\chi,j} > 0$, 
and positive integer powers of $\Phi^\dagger$ when $f_{\psi,i} + f_{\chi,j} < 0$.

Once $\Phi$ acquires a VEV,
the $\UFN$-symmetric terms in Eq.\,\eqref{eq:LY} give rise to  the SM Yukawa interactions, 
\begin{gather}
\label{eq:hadamard}
   y^{(u)}_{ij} = \kappa^{(u)}_{ij}\, \epsilon^{q^{(u)}_{ij}}\ , 
\quad
   y^{(d)}_{ij} = \kappa^{(d)}_{ij}\,  \epsilon^{q^{(d)}_{ij}}\ ,
\quad
   y^{(e)}_{ij} = \kappa^{(e)}_{ij}\,  \epsilon^{q^{(e)}_{ij}}\ .
\end{gather}
Here, 
summation over $i,j$ is not implied in Eq.\,\eqref{eq:hadamard}.%
\footnote{When all $f_{i}+f_{j}$ have the same sign, the term $\kappa_{ij}\, \epsilon^{f_{i}+f_{j}}$ can be expressed as a matrix multiplication:
$
\kappa_{ij}\, \epsilon^{f_i+f_j} = (\epsilon^{f} \cdot \kappa \cdot \epsilon^{f})_{ij},
$
where $\epsilon^{f}_{ij} = \epsilon^{f_i} \delta_{ij}$. }
The FN breaking parameter,
$\epsilon$, 
and the exponents $q_{ij}$ are defined as
\begin{gather}
\label{eq:delta}
   \epsilon \equiv \frac{\langle\Phi\rangle}{M_{*}} = \frac{\langle\Phi^{\dagger}\rangle}{M_{*}}\ , \\
\label{eq:chargeq}
   q^{(u)}_{ij} = |f_{Q,i} + f_{\ubar,j}|\ , 
\quad
   q^{(d)}_{ij} = |f_{Q,i} + f_{\dbar,j}|\ ,
    \quad
   q^{(e)}_{ij} = |f_{L,i} + f_{\ebar,j}|\ .
\end{gather}
Here, we have taken $\epsilon>0$ without loss of generality.
Since we assume $\epsilon = \order{0.1}$,
the above couplings are more suppressed for larger values of $q_{ij}$,
which generates non-trivial hierarchies of the Yukawa couplings even for $\kappa$'s of $\order{1}$.

The FN mechanism also affects the masses and mixing structure of neutrinos.
The Yukawa coupling constants of the neutrino sector are given by
\begin{gather}
\label{eq:hadamard_nu}    
   y^{(D)}_{ia} = \kappa^{(D)}_{ia}\, \epsilon^{q_{ia}^{(D)}}\ ,
   \quad
   y^{(R)}_{ab} = \kappa^{(R)}_{ab}\, \epsilon^{q_{ab}^{(R)}}\ , \\
   q^{(D)}_{ia} = |f_{L,i}+f_{\Nbar,a}|\ ,
   \quad 
   q^{(R)}_{ab} = |f_{\Nbar,a}+f_{\Nbar,b}|\ .
\end{gather}
Here, 
$\kappa^{(D)}$ is 
a complex $3\times 3$ matrix,
while $\kappa^{(R)}$
is a complex $3\times 3$ symmetric matrix.
As before, summation over $i, a, b$ is not implied in Eq.\,\eqref{eq:hadamard_nu}.

In the conventional SU(5) GUT, 
the FN charges of the 
quarks and leptons satisfy the relations,
\begin{gather}
f_{Q,i} = f_{\ubar,i} = f_{\ebar,i}=f_{\ten,i} \ , \quad
f_{\dbar,i} = f_{L,i}=f_{\fivebar,i}\ ,
\end{gather}
since they are embedded within $\ten$ and $\fivebar$ representations.
Due to the $\UFN$ symmetry, the 
Yukawa interactions in the GUT (similar to those described by Eq.\,\eqref{eq:quark Yukawa}) are modified in a manner analogous to  Eq.\,\eqref{eq:LY}.
The FN charges of the right-handed neutrinos, however, can be chosen independently.
In the minimal supersymmetric SU(5) GUT model,
for example,
the following charge assignments were proposed in Ref.\,\cite{Sato:1997hv},
\begin{align}
\epsilon \sim 0.06\ , \quad
f_{\ten} = (2,1,0)\ , \quad
f_{\fivebar} = (a+1, a, a)\ , \quad
f_{\Nbar} = (d,c,b)\ , \quad
\end{align}
where $a=0$ or $1$ and $0\le b \le c \le d$
(see also Ref.\,\cite{Babu:2016aro}).

In the SU(5)$\times$U(2)$_H$ fake GUT model, 
since the quarks originate from the SU(5) multiplets,
the effective Yukawa couplings in the quark sector 
are determined by 
the FN charges of $\fivebar_i$ and $\ten_i$,
similar to the conventional GUT.
The Yukawa interactions for charged leptons,
on the other hand, 
have two UV origins as shown in Eq.\,\eqref{eq:Yukawa_Origins}.
However, 
as mentioned earlier, 
lepton mixing angles
in Eqs.\,\eqref{eq:L_mixing}
and \eqref{eq:E_mixing}
are highly suppressed to avoid the constraints from the proton decay.
Accordingly, 
the SM leptons almost entirely originate from $L_H$ and $\Ebar_H$
and the SM lepton Yukawa coupling
is effectively given by $y^{(LE)}$,
regardless of the details of the lepton mixing angles, $\theta_{L,E}$.
Accordingly,
the flavor structure
of the charged lepton Yukawa coupling 
is determined by the FN charges 
of $L_{Hi}$ and $\Ebar_{Hi}$.
Besides, since 
the right-handed neutrino Yukawa 
coupling is given by Eq.\,\eqref{eq:all_nbar_int},
the flavor structure of the neutrino Yukawa couplings is determined by the FN charges of $L_{Hi}$ and $\Nbar_a$.

Altogether, 
the relevant FN charges in the SU(5)$\times$U(2)$_H$ fake GUT model are
\begin{align}
f_{\ten}\ , \quad 
f_{\fivebar}\ , \quad 
f_{L_H} \ , \quad f_{\Ebar_H}\ ,
\quad f_{\Nbar} \ .
\end{align}
Effective Yukawa couplings are obtained by using these FN charges,
\begin{gather}
\label{eq:effective_yukawa_fake}    
   y^{(10,5,LE,LN,R)}_{ij} = \kappa^{(10,5,LE,LN,R)}_{ij}\, \epsilon^{q_{ij}^{(10,5,LE,LN,R)}}\ , \\
   q^{(10)}_{ij} = |f_{\ten,i}+f_{\ten,j}|\ ,
   \quad 
   q^{(5)}_{ij} = |f_{\ten,i}+f_{\fivebar,j}|\ , \\
   q^{(LE)}_{ij} = |f_{L_H,i}+f_{\Ebar_H,j}|\ , \quad
   q^{(LN)}_{ij} = |f_{L_H,i}+f_{\Nbar,j}|\ , \quad
   q^{(R)}_{ij} = |f_{\Nbar,i}+f_{\Nbar,j}|\ .
\end{gather}
Note that the FN charge assignment of $\Lbar_{Hi}$ and $E_{Hi}$ do not 
affect the flavor structure of the leptons as long as $|\theta_{L,E}|\ll 1$.
In the next section, 
we show examples of the FN charge assignment.

\subsection{Benchmark FN charge assignment in the 
\texorpdfstring{$\boldsymbol{\SU(5) \times \mathrm{U}(2)_{H}}$}{} model}
\label{sec:good_charge}

In our analysis, 
we compare the SM parameters predicted by the FN mechanism with those evaluated in the $\overline{\mathrm{MS}}$ scheme at the renormalization scale $\mu_R = 10^{14}$\,GeV. 
We use the PDG averages of the SM particle masses \cite{ParticleDataGroup:2022pth} as input parameters,
and relevant renormalization group equations (RGEs) in 
the $\overline{\mathrm{MS}}$\,\cite{Buttazzo:2013uya}.
We estimate the Yukawa couplings of the light quarks, 
by using the QCD four-loop RGEs and three-loop decoupling effects from heavy quarks\,\cite{Chetyrkin:1997sg}.
Note that the discussion is not highly sensitive to the choice of the renormalization scale, as long as it is significantly higher than the electroweak scale. 
This is because the dominant renormalization group effects stem from QCD interactions below the electroweak scale. 
For this reason, 
we also neglect the running effects between $\mu_R$ and $M_*$.

For the neutrino mass parameters,
we do not account for the renormalization group effects,
since the relative sizes of the FN breaking scale 
and $M_R$ are not fixed 
in this work.
Moreover, 
the renormalization group effects
in the SM interaction 
have a negligible impact on the neutrino mass ratio and mixing angles, 
which are utilized in the following analysis.

\begin{figure}[t]
	\centering
 	\subcaptionbox{BP1}
{\includegraphics[width=0.49\textwidth]{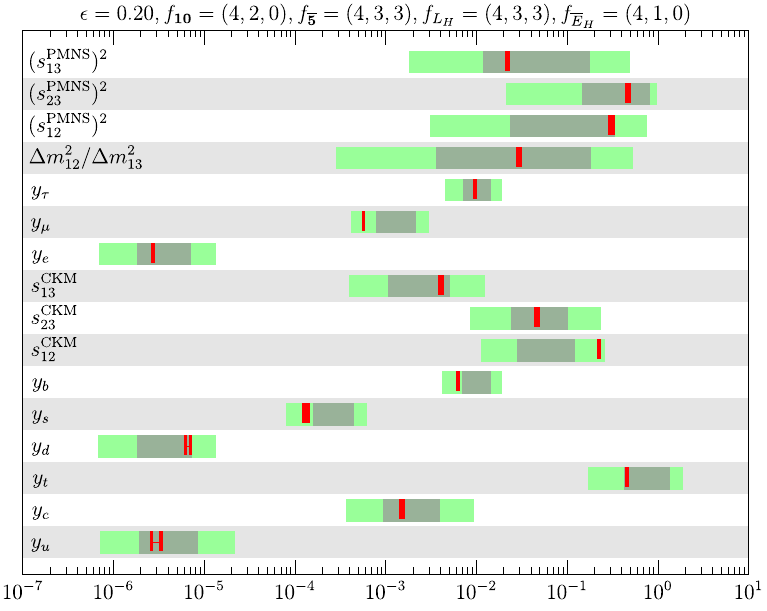}}
 	\subcaptionbox{BP2}
{\includegraphics[width=0.49\textwidth]{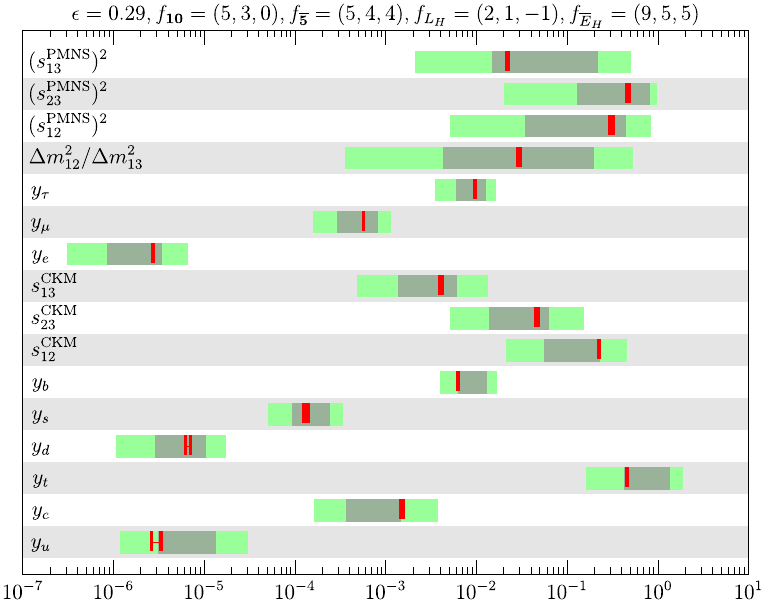}}
\caption{
The prior distributions for the SM parameters 
from $\order{1}$ distributions of $\kappa$'s with the FN charge assignments 
in Eqs.\,\eqref{eq:pattern_a}
and \eqref{eq:pattern_b}.
The red bars represent the values of the running parameters estimated within the $\overline{\mathrm{MS}}$ scheme at $\mu_{R}=10^{14}$ GeV.
The observed values are within the ranges of the posterior distributions.
}
\label{fig:fakeFN}
\end{figure}

In this paper, 
we take two benchmark points for the FN charge assignments
of the fake GUT multiplets,
\begin{align}
\label{eq:pattern_a}
&\mathrm{BP1} : \quad
\epsilon = 0.20\ , \quad
f_{\ten} = (4,2,0) \ , \quad 
f_{\fivebar} = (4,3,3)\ , \quad 
f_{L_H} = (4,3,3)\ , \quad
f_{\Ebar_H} =(4,1,0)\ , \\
\label{eq:pattern_b}
&\mathrm{BP2} : \quad
\epsilon = 0.29\ , \quad
f_{\ten} = (5,3,0) \ , \quad 
f_{\fivebar} = (5,4,4)\ , \quad 
f_{L_H} = (2,1,-1)\ , \quad
f_{\Ebar_H} =(9,5,5)\ .
\end{align}
Here, 
we set all $f_{\Nbar,a}=0$ ($a=1,2,3$) and $\sin \theta_h = \cos \theta_h = 1/\sqrt{2}$ for simplicity.
Additionally, 
we assume $H_5$ and $H_2$ do not have any FN charges.

In Fig.\,\ref{fig:fakeFN},
we present the distributions of the 
predicted parameters for these FN charge assignments
with random $\order{1}$ coefficients $\kappa$’s.
We assume the normal neutrino mass ordering, 
i.e., ($m_1 < m_2 < m_3$).
The dark green and light green bands represent the $1\sigma$ and $2\sigma$ percentiles, respectively.
The red bars indicate the values of the running parameters which are estimated in the $\overline{\mathrm{MS}}$ scheme at $\mu_R=10^{14}$ GeV.
From this figure, 
we find that 
the observed physical parameters are within typical ranges of 
the predictions of the FN mechanism with
the above FN charge assignments.

In the above analysis, 
we adopt the following distributions 
for the $\order{1}$ coefficients 
$\kappa^{(10,5,LE,LN,R)}$ in Eq.\,\eqref{eq:effective_yukawa_fake}.
For the Dirac-type couplings $\kappa^{(10,5,LE,LN)}_{ij}$,
we use the following distributions,
\begin{align}
\label{eq:distribution_1}
\Re\, \kappa^{(10,5,LE,LN)}_{ij}
= \frac{1}{\sqrt{2}}\, \calN(0,1)\ ,
\quad \Im\, \kappa^{(10,5,LE,LN)}_{ij}
= \frac{1}{\sqrt{2}}\, \calN(0,1)\ .
\end{align}
Here, 
$\calN(0,1)$ denotes a random number drawn from a normal distribution with a mean of $0$ and a standard deviation of $1$.
Similarly, 
for the Majorana-type couplings $\kappa^{(R)}_{ij}$,
we assume the following distributions,
\begin{gather}
\Re\, \kappa^{(R)}_{aa} 
= 
\frac{1}{\sqrt{2}}\, \calN(0,1)\, , \quad
\Im\, \kappa^{(R)}_{aa}
= 
\frac{1}{\sqrt{2}}\, \calN(0,1)\, , \quad \nonumber \\
\Re\, \kappa^{(R)}_{ab} 
= \frac{1}{2}\, \calN(0,1)\,  , \quad
\Im\, \kappa^{(R)}_{ab}
= \frac{1}{2}\, \calN(0,1)\, ,\quad (a> b)\ .
\label{eq:distribution_2}
\end{gather}
Note that we assume the variance of the off-diagonal elements is
smaller than that of the diagonal elements by a factor of $\sqrt{2}$,
which makes the distributions flavor symmetric.

\section{Nucleon Decay with U(1) Flavor Symmetry}
\label{sec:fake_flavor}

As mentioned earlier, 
the generation of the leptonic components in $L_{\fivebar, i}$ and $\Ebar_{\ten, i}$ within the SU(5) sector does not align with 
that of the SM leptons in the mass eigenstates. 
In this section, 
we discuss how the FN charge assignments affect 
the flavor dependence 
of the leptonic components 
in $L_{\fivebar, i}$ and $\Ebar_{\ten, i}$,
which provides striking effects 
on the nucleon decay rates and the branching fractions.

\subsection{Flavor effects on lepton mixing angles}
\label{sec:fake_flavor_concrete_model}

Let us discuss the effects of the FN mechanism on the lepton mixing terms in  Eq.\,\eqref{eq:Abelian_mass}.
Under the FN mechanism, they are modified as,
\begin{align}
\label{eq:U(2)_massFN}
\mathcal{L} 
&= 
\overline{m}_L\, \kappa^{(m_L)}_{ij}\,
\epsilon^{f_{\Lbar_{H},i}+f_{L_{H},j}}\,
\Lbar_{Hi}\, L_{Hj}\, 
+ 
\kappa^{(\lambda_L)}_{ij}\, 
\epsilon^{f_{\Lbar_{H},i} + f_{\fivebar,j}} \,
\Lbar_{Hi}\, \phi_{2}\, \fivebar_j \nonumber \\ 
&+ 
\overline{m}_E\,
\kappa^{(m_E)}_{ij}\,
\epsilon^{f_{E_{H},i} + f_{\Ebar_{H},j}}\,
E_{Hi}\, \Ebar_{Hj}\,
+ 
\frac{\kappa^{(\lambda_E)}_{ij}}{\Lambda_{\mathrm{cut}}}\, 
\epsilon^{f_{E_{H},i} + f_{\ten,j}}\,
E_{Hi}\, \phi_2^\dagger\, \phi_2^\dagger\, \ten_j
+ h.c.
\end{align}
Here, 
$\kappa$'s are dimensionless $\order{1}$ coefficients that follow the Gaussian distribution 
for Dirac-type couplings described in the previous section.
The parameters 
$\overline{m}_L$ and $\overline{m}_E$ 
are dimensionful constants.
We assume $\phi_2$ does not have the FN charge.
We also focus on the case 
where each sums of the FN charges in Eq.\,\eqref{eq:U(2)_massFN} are larger than or equal to $0$.
In subsequent calculations,
we choose the values for $\overline{m}_L$, $\overline{m}_E$ and $\Lambda_{\mathrm{cut}}$ to achieve the tiny lepton mixing angles in Eqs.\,\eqref{eq:L_mixing} 
and \eqref{eq:E_mixing}.
The relationships between the coefficients of Eqs.\,\eqref{eq:Abelian_mass} and \eqref{eq:U(2)_massFN} are given by
\begin{align}
m_{L,ij} 
&= \overline{m}_L\, \kappa^{(m_L)}_{ij} 
\epsilon^{f_{\Lbar_{H},i}+f_{L_{H},j}}\ , \quad
\lambda_{L,ij}
= \kappa^{(\lambda_L)}_{ij} 
\epsilon^{f_{\Lbar_{H},i} + f_{\fivebar,j}}\ , \nonumber \\
m_{E,ij}
&= \overline{m}_E\, \kappa^{(m_E)}_{ij} 
\epsilon^{f_{E_{H},i} + f_{\Ebar_{H},j}}\ , \quad
\lambda_{E,ij}
= 
\kappa^{(\lambda_E)}_{ij}
\epsilon^{f_{E_{H},i} + f_{\ten,j}}\ .
\end{align}
Note that the summation over the flavor indices $i,j=1,2,3$ is not taken.

Let us consider the mass terms of the doublet leptons in Eq.\,\eqref{eq:leptonic_mass},
\begin{align}
\calL_{\mathrm{mass}}
&= \Lbar_{Hi}\,
\begin{pmatrix}
\lambda_{L,ij}\, v_{2} & m_{L,ik}
\end{pmatrix}
\begin{pmatrix}
L_{\fivebar, j} \\
L_{Hk}
\end{pmatrix} \nonumber \\
&= (\Lbar_{Hi}\, \lambda_{L,ij}\, v_{2})
(L_{\fivebar, j} + v_{2}^{-1}\, (\lambda_{L}^{-1})_{jn}\, m_{L,nk}\, L_{Hk})\ .
\end{align}
From this expression, 
the heavy leptons $L_{Mi}$ and the SM leptons $\ell_{\SM,i}$ are given by,
\begin{align}
L_{Mi}
&\simeq L_{\fivebar, i} 
+ v_{2}^{-1}\, (\lambda_{L}^{-1})_{ik}\, m_{L,kj}\, L_{Hj}\ , \\
\ell_{\SM,i}
&\simeq L_{Hi}
- v_{2}^{-1}\, (\lambda_{L}^{-1})_{ik}\, m_{L,kj}\, L_{\fivebar, j}
\ .
\end{align}
Thus, for small lepton mixing angles $\theta_L$,  
we find that the 
$L_{\fivebar}$'s include the SM
lepton contents as,
\begin{align}
L_{\fivebar, i}
\simeq
L_{Mi} - v_2^{-1} (\lambda_L^{-1})_{ik}\, m_{L,kj}\, \ell_{\SM,j}
\ .
\end{align}

Furthermore,
for the case where 
each sum of the FN charges in Eq.\,\eqref{eq:U(2)_massFN} 
is larger than or equal to $0$,
as we are considering,
$\lambda_L^{-1}\,m_L$ does not depend on the FN charge of $\Lbar_H$.
Let us see this explicitly.
In this case, 
$\lambda_L$ and $m_L$ can be expressed as products of matrices,
\begin{align}
\lambda_L 
= \epsilon^{f_{\Lbar_H}}\,\kappa^{(\lambda_L)}\,\epsilon^{f_{\fivebar}}\ , \quad
m_L
= \overline{m}_L\, \epsilon^{f_{\Lbar_H}}\,\kappa^{(m_L)}\,\epsilon^{f_{L_H}}\ ,
\end{align}
where $\epsilon^{f}_{ij} = \epsilon^{f_i} \delta_{ij}$.
Therefore,
$\lambda_L^{-1}\,m_L$ is given as
\begin{align}
(\lambda_L)^{-1}\,m_L 
= \overline{m}_L\,(\epsilon^{f_{\fivebar}})^{-1}\,(\kappa^{(\lambda_L)})^{-1}\,
(\epsilon^{f_{\Lbar_H}})^{-1}\,
\epsilon^{f_{\Lbar_H}}\,
\kappa^{(m_L)}\,
\epsilon^{f_{L_H}}
= \overline{m}_L\,(\epsilon^{f_{\fivebar}})^{-1}\,(\kappa^{(\lambda_L)})^{-1}\,
\kappa^{(m_L)}\,
\epsilon^{f_{L_H}}\ .
\end{align}
This shows that the FN charge matrix of
$\Lbar_H$ cancels out in $\lambda_L^{-1}\,m_L$.
As a result,
the SM lepton components in the $L_{\fivebar}$
are independent of the FN charges of $\Lbar_{Hi}$,
since they
are evaluated as 
\begin{align}
\label{eq:FNangle}
(\theta_L)_{ij}=v_2^{-1}(\lambda_{L}^{-1})_{ik}\, m_{L,kj} 
\sim 
\overline{\theta}_L \times\epsilon^{-f_{\fivebar,i} + f_{L_{H},j}}
\ .
\end{align}
The same procedure applies to the case of the singlet leptons $\Ebar_H$,
\begin{align}
\label{eq:FNangleE}
(\theta_E)_{ij}=v_2^{-2}\, \Lambda_{\mathrm{cut}}\,(\lambda_{E}^{-1})_{ik}\, m_{E,kj} 
&\sim \overline{\theta}_E \times\epsilon^{-f_{\ten,i} + f_{\Ebar_{H},j}} \ .
\end{align}
For later purpose, 
we have defined
\begin{align}
\label{eq:thetabar}
\overline{\theta}_L 
= \frac{\overline{m}_L}{v_2}\ , \quad
\overline{\theta}_E 
= \frac{\overline{m}_E\, \Lambda_{\mathrm{cut}}}{v_2^2}\ .
\end{align}

A notable feature of the FN mechanism in the fake GUT is that
SM lepton contents within the SU(5) multiplets 
in Eq.\,\eqref{eq:FNangle}
tend to be more suppressed 
for the lower generations.
This is because 
the $L_H$ and $\Ebar_H$ of the lower generations has larger FN charges, 
and the generations of these vector-like leptons correspond to those of the SM leptons.
Therefore,
nucleons are more likely to decay into the SM leptons of the higher generations.

As discussed in Sec.\,\ref{sec:Nucleondecay_review},the lepton mixing angles $\theta_{L,E}$ must be constrained to be below $\order{10^{-4}}$ to avoid violating nucleon lifetime constraints. 
For the two benchmark points, 
BP1 and BP2, 
this requires the U(1)$_{LH}$ breaking parameters to be on the order of $\overline{m}_{L,E} \simeq 10^{8\mathchar`-9}$\,GeV, 
given that $\Lambda_{\mathrm{cut}} \gtrsim v_2$. 
This scale is reasonably close to the right-handed neutrino mass scale $M_R$, 
which is relevant for the seesaw mechanism under the assumptions for these benchmark points. 
Although we do not explore this further in the present work, 
this observation suggests that the origin of the U$(1)_{LH}$ symmetry-breaking parameters $\overline{m}_L$, $\overline{m}_E$, and $M_R$ may have a common origin in a unified sector.

\subsection{Flavorful nucleon decay}

Now,
let us discuss the predictions of nucleon decay.
For given values of $\kappa$,
we first transform the random flavor basis of 
SU(5) multiplets 
to a basis where 
the up-type Yukawa matrix is diagonal,
and 
the down-type Yukawa matrix is given by
the diagonal matrix multiplied by the CKM matrix,
as described in Ref.\,\cite{Nagata:2013sba}.
We also extract the SM lepton components,
which correspond to the massless eigenstates of the lepton mass matrices in Eq.\,\eqref{eq:leptonic_mass_matrix}).
Subsequently, 
we transform the SM leptons to their mass basis.
In this way, 
we obtain the lepton components in the mass basis
of the SU(5) multiplets in the down-type CKM basis as
\begin{align}
\label{eq:mixing angle matrix}
L_{\fivebar, i} 
\supset (\theta_L)_{ij}\, \ell_{\SM,j}\ , \quad
\Ebar_{\ten, i} 
\supset (\theta_{E})_{ij}\,\ebar_{\SM,j}\ ,
\end{align}
for each realization.
Note that $\theta_L$ and $\theta_{\ebar}$ are $3 \times 3$ matrices,
and 
do not depend on either $f_{\Lbar_{H}}$'s or $f_{E_{H}}$'s.

Once we obtain these matrices, 
we can calculate the nucleon decay rate using the following
operators,
\begin{align}
\label{eq:O6}
\mathcal{O}_{ijkl}^{(1)} 
= 
\epsilon_{abc} \left(\ubar_i^{\dagger a}\, \dbar_j^{\dagger b} \right) 
\left(Q_k^c\, L_{\fivebar,l} \right)\ , \quad
\mathcal{O}_{ijkl}^{(2)} 
= 
\epsilon_{abc} \left(Q_i^a\, Q_j^b \right) \left(\ubar_k^{\dagger c}\, \Ebar^{\dagger}_{\ten,l} \right)\ ,
\end{align}
which are generated by the heavy SU(5) gauge bosons (see Fig.\,\ref{fig:tikz_dim6}).
The Wilson coefficients for these operator are given by
\begin{align}
C^{ijkl}_{(1)}
= - \frac{g_5^2}{M_X^2}\, e^{i \varphi_i}\, \delta^{ik}\, \delta^{jl}\ , \quad
C^{ijkl}_{(2)}
= - \frac{g_5^2}{M_X^2}\, e^{i \varphi_k}\, \delta^{ik}\, \delta^{jl}\ .
\end{align}
Here,
the phases, $\varphi_i$, satisfy $\sum_i  \varphi_i = 0$,
and 
arise from the basis transformation explained earlier.
These operators are given at the fake GUT scale.
In order to get low-energy observable predictions,
we  
solve the RGE equations of these Wilson coefficients following Refs.\,\cite{Abbott:1980zj,Nihei:1994tx}.

In our analysis, 
we utilize matrix elements for proton decay estimated by the QCD lattice calculation
in Ref.\,\cite{Yoo:2021gql}.
For $\eta$ meson modes, 
we use the results in Ref.\,\cite{Aoki:2017puj}.
Additionally, 
we obtain the neutron matrix elements 
for baryon number-violating neutron decays
through the SU(2) isospin symmetry of QCD matrix elements.
Therefore,
the neutron decay rate into a pion has correlation with the proton decay rate into a pion as follows,
\begin{align}
\label{eq:pn_correlation}
\Gamma(n \to \pi^- \ell^+_i)
&= 2\, \Gamma(p \to \pi^0 \ell^+_i)\ , \\
\Gamma(n \to \pi^0 \overline{\nu})
&= \frac{1}{2}\, \Gamma(p \to \pi^+ \overline{\nu})\ .
\end{align}
Moreover,
while the proton can decay into modes including a charged kaon,
the neutron cannot decay into such modes within this model.
This follows from the following Gell-Mann-Nishijima formula,
which describes the relationship between various quantum numbers of a particle:
\begin{align}
Q = I_3 + \frac{1}{2}\,(B+S)\ ,
\end{align}
where $Q,I_3,B$ and $S$ denote the electric charge, 
the third component of the SU(2) isospin, 
the baryon number and the strangeness of a particle, respectively.
Since electric charge is conserved,
the decay
$n \to K^- \ell^+_i$
is forbidden\,\cite{Machacek:1979tx,Weinberg:1979sa}.

To obtain predictions for nucleon decays, 
we follow the Monte-Carlo procedure outlined below. 
For given sets of the FN charges and the FN breaking parameter $\epsilon$ in Eqs.\,\eqref{eq:pattern_a} and \eqref{eq:pattern_b},
we first generate all the $\kappa$ values by sampling from Gaussian distributions in Eqs.\,\eqref{eq:distribution_1} and \eqref{eq:distribution_2}.
We obtain posterior distributions of $\kappa$'s which take into the observed parameters (the masses and mixing angles of the SM quarks and SM leptons) in Fig.\,\ref{fig:fakeFN}.
From those posterior distributions,
we predict
the nucleon decay rates 
for each decay mode.
We also incorporate uncertainties from the QCD matrix elements in the predictions.

\subsection{Results}

\begin{figure}[t!]
\centering

\raisebox{50pt}{\subcaptionbox{Contribution of $\mathcal{O}^{(1)} (\mathrm{or}\,\, \overline{\theta}_E = 0)$ for BP1.
\label{fig:BP1L}}
{\includegraphics[width=0.49\textwidth]{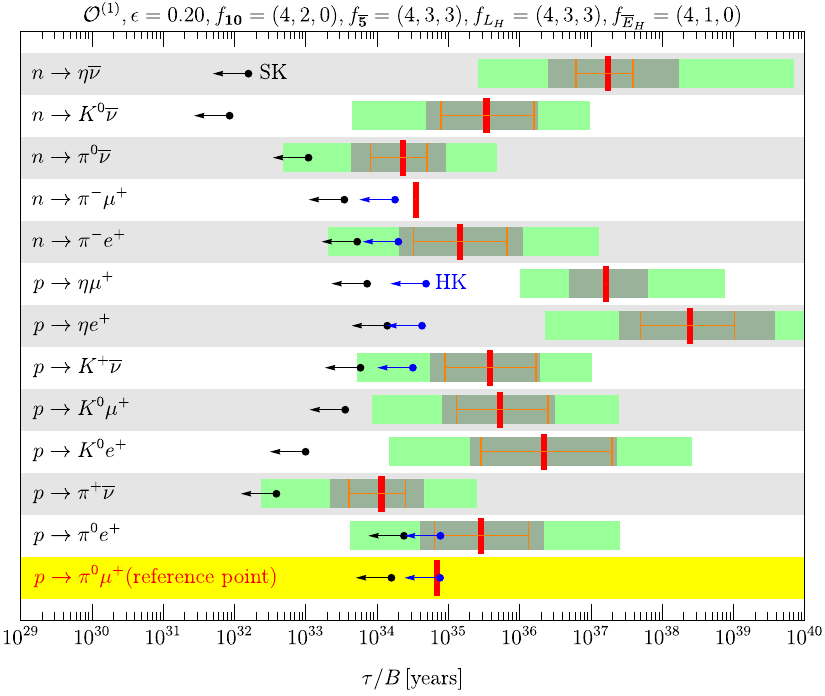}}}
\raisebox{50pt}{
\subcaptionbox{Contribution of $\mathcal{O}^{(2)}(\mathrm{or}\,\, \overline{\theta}_L = 0)$ for BP1.
\label{fig:BP1R}}
{\includegraphics[width=0.49\textwidth]{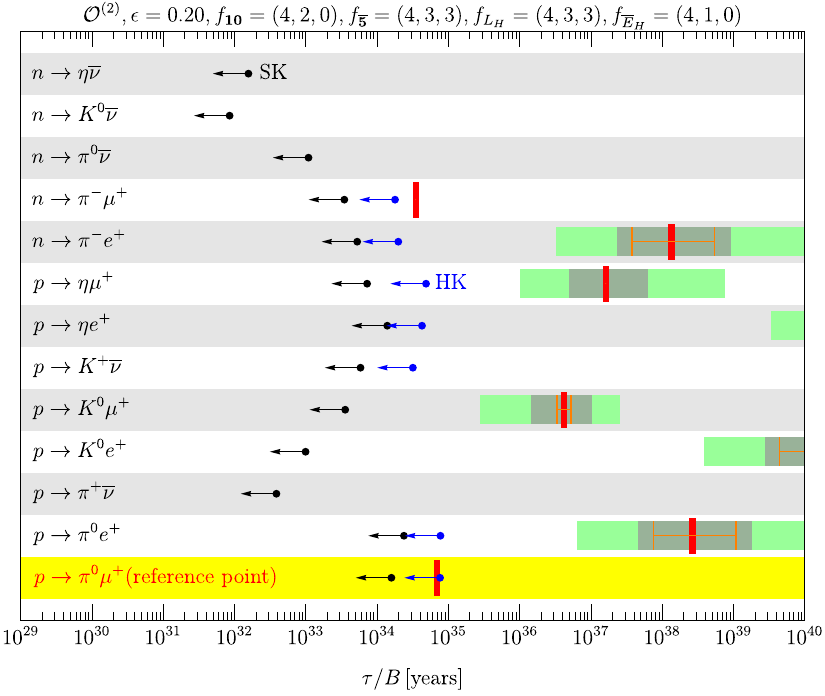}}}
\subcaptionbox{Contribution of $\mathcal{O}^{(1)}(\mathrm{or}\,\, \overline{\theta}_E = 0)$ for BP2.
\label{fig:BP2L}}
{\includegraphics[width=0.49\textwidth]{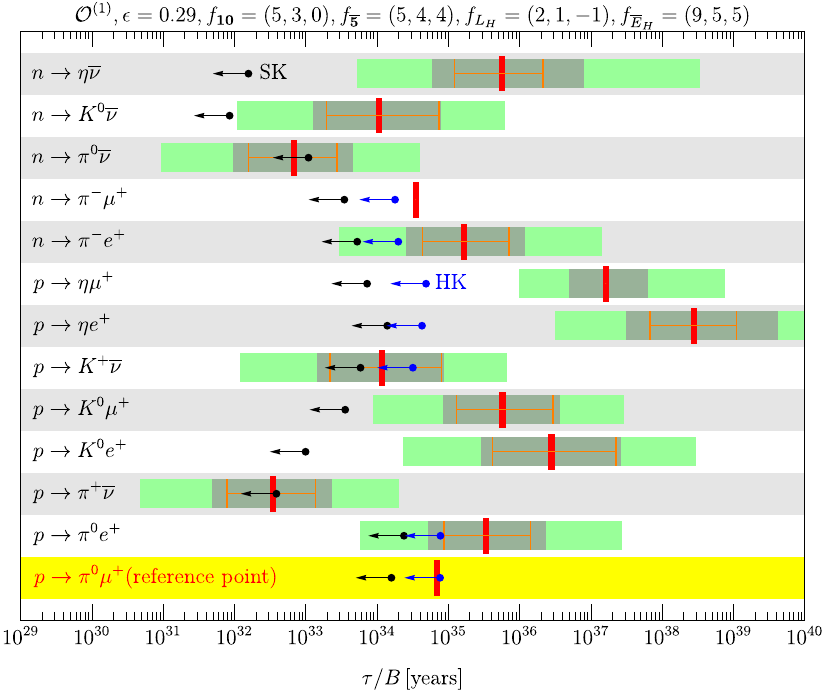}}
\subcaptionbox{Contribution of $\mathcal{O}^{(2)}(\mathrm{or}\,\, \overline{\theta}_L = 0)$ for BP2.
\label{fig:BP2R}}
{\includegraphics[width=0.49\textwidth]{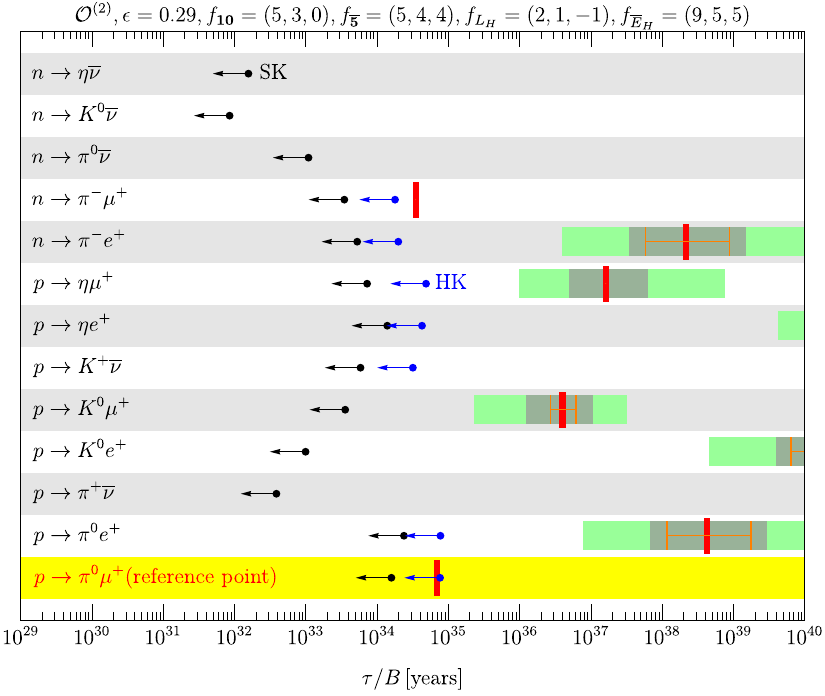}}
\caption{The nucleon lifetimes for various decay modes in benchmark points BP1 and BP2. 
Figures (a) and (b) show the nucleon decay originating from the operators $\mathcal{O}^{(1)}$ and $\mathcal{O}^{(2)}$ in Eq.\,\eqref{eq:O6} for BP1,
respectively. 
Figures (c) and (d) illustrate the same for BP2.
We set $M_X = 10^{14}$\,GeV and 
adjust $\overline{\theta}_L$ and 
$\overline{\theta}_E$ to 
achieve a reference nucleon lifetime for the $p \to \pi^0 \mu^+$ decay mode of $7.0 \times 10^{34}$ years. 
The dark green and light green bands represent the $1 \sigma$ and $2 \sigma$ percentile ranges of the lifetime distributions, with median values indicated by red bars. 
Orange bars depict the $1 \sigma$ percentile range excluding uncertainties in QCD matrix elements.
Current constraints from SK are shown with black arrows\,\cite{Super-Kamiokande:2020wjk,Super-Kamiokande:2013rwg,Super-Kamiokande:2005lev,Super-Kamiokande:2022egr,Super-Kamiokande:2014otb,Super-Kamiokande:2024qbv}, 
and future projections for HK are indicated with blue arrows\,\cite{Hyper-Kamiokande:2018ofw}.
}
\label{fig:decayBP1}
\end{figure}

The contributions of $\mathcal{O}^{(1)}$ and $\mathcal{O}^{(2)}$ to the nucleon decay amplitudes are suppressed by
$\overline{\theta}_L$
and $\overline{\theta}_E$,
respectively.
We therefore treat these contributions independently. 
As we will demonstrate,
$p \to \pi^0 \mu^+$ mode is one of the most promising discovery channels in the present model,
and hence we choose this decay mode as a reference point.
In the following analysis,
we fix $M_X = 10^{14}$\,GeV, 
and adjust either $\overline{\theta}_L$ or $\overline{\theta}_E$ such that
the lifetime of the $p \to \pi^0 \mu^+$ mode is $7.0 \times 10^{34}$\,years for each realization of $\kappa$ values.
The distributions of other decay modes are provided relative to this fixed lifetime for the $p \to \pi^0 \mu^+$ mode.

In Fig.\,\ref{fig:decayBP1}, 
we present the percentile ranges of the posterior distributions for the predicted nucleon lifetimes across various decay modes for the FN charge assignments of BP1 and BP2.
In Figs.\,\ref{fig:BP1L} and \ref{fig:BP1R}  (Figs.\,\ref{fig:BP2L} and \ref{fig:BP2R}), 
we illustrate the nucleon lifetimes arising from the nucleon decay operators $\mathcal{O}^{(1)}$ ($\mathcal{O}^{(2)}$) in Eq.\,\eqref{eq:O6}.
For decay modes involving $\overline{\nu}$,
we sum over the contributions from all neutrino flavors. 
The dark green and light green bands represent the $1 \sigma$ and $2 \sigma$ percentile ranges of the distributions, respectively, 
with red bars indicating the medians. 
The orange bars represent the $1 \sigma$ percentile range arising from uncertainties in the $\order{1}$ coefficients, 
$\kappa$, 
while excluding uncertainties in the QCD matrix elements.
This uncertainty typically affects the predicted lifetime by a factor of $\order{10}$. 
Note that the ratio $\tau(n \to \pi^- \mu^+)/\tau(p \to \pi^0 \mu^+)$ is free from uncertainty due to the SU(2) isospin symmetry in Eq.\,\eqref{eq:pn_correlation}.
The uncertainty in the ratio $\tau(p \to \eta \mu^+)/\tau(p \to \pi^0 \mu^+)$ is primarily driven by the uncertainty in the QCD matrix element, 
with no contribution from the $\kappa$'s since these decays are caused by the common effective operators at the QCD scale.
Black and blue arrows indicate current constraints from SK\,\cite{Super-Kamiokande:2020wjk,Super-Kamiokande:2013rwg,Super-Kamiokande:2005lev,Super-Kamiokande:2022egr,Super-Kamiokande:2014otb,Super-Kamiokande:2024qbv} and the future projections for HK\,\cite{Hyper-Kamiokande:2018ofw}.

These figures highlight a clear contrast with the minimal SU(5) GUT. 
In the minimal SU(5) GUT, 
the dominant decay mode is $p \to \pi^0 e^+$, 
with other lepton flavor modes being significantly suppressed. 
In contrast, 
within the fake GUT, 
the generations of quarks and leptons are not aligned, 
allowing for a broader spectrum of decay modes. 
Specifically, 
under the FN mechanism, 
each SU(5) multiplet includes substantial contributions from the second and third generations of SM leptons. 
Consequently, 
nucleon decays into $\mu^+$ and $\overline{\nu}_{\mu,\tau}$ are more likely. 
The figures demonstrate that the $p \to \pi^0 \mu^+$ mode emerges as a leading discovery channel. 
This distinctive feature of the FN mechanism within the fake GUT is clearly depicted in the figures for both BP1 and BP2. 
Additionally, 
it should be noted that the FN mechanism does not alter nucleon decay modes mediated by GUT gauge boson exchange in the minimal GUT. 
Therefore, 
detecting various lepton flavor modes would be highly effective for probing the fake GUT.

The notable difference between Figs.\,\ref{fig:BP1L} and \ref{fig:BP1R} arises because $\mathcal{O}^{(1)}$ includes a doublet lepton, 
while $\mathcal{O}^{(2)}$ contains a singlet lepton. 
This difference results in the absence of neutrino modes in Fig.\,\ref{fig:BP1R}, 
which are present in Fig.\,\ref{fig:BP1L}. 
A similar pattern is observed between Fig.\,\ref{fig:BP2L} and Fig.\,\ref{fig:BP2R}. 
In contrast, 
in the minimal SU(5) GUT,
the contribution from $\mathcal{O}^{(2)}$ to $\Gamma(p \to \pi^0 e^+)$ is 
approximately four times larger than that from $\mathcal{O}^{(1)}$.
Consequently, 
the search for neutrino decay modes serves as an essential probe for distinguishing the fake GUT.

As we can see in Eq.\,\eqref{eq:FNangle}, 
the difference in $\tau(p \to \pi^0 e^+)/ \tau(p \to \pi^0 \mu^+)$ between Figs.\,\ref{fig:BP1L} and \ref{fig:BP1R} primarily stems from the variations in $f_{L_{H},2}-f_{L_{H},1}$ and $f_{\overline{E}_{H},2}-f_{\overline{E}_{H},1}$. 
The greater this charge difference, 
the larger the ratio $\tau(p \to \pi^0 e^+)/ \tau(p \to \pi^0 \mu^+)$ becomes. 
This behavior also applies to the baryon-number violating neutron decay. 
This result contrasts with conventional GUT predictions, where the ratio is $\tau(p \to \pi^0 e^+)/\tau(p \to \pi^0 \mu^+) = \order{10^{-(2\mathchar`-3)}}$.

The figures illustrate that the predicted branching fractions are influenced by the FN charges. For example, decays into neutrino modes in Fig.\,\ref{fig:BP2L} proceed more rapidly than those in Fig.\,\ref{fig:BP1L}. This difference arises from the ratios of the lepton mixing angles,
\begin{align}
\label{eq:three_lepton_mixing}
\frac{(\theta_L)_{i1}^{(\mathrm{BP2})}}{(\theta_L)_{i1}^{(\mathrm{BP1})}}   \sim \epsilon^{-3}\ , \quad
\frac{(\theta_L)_{i2}^{(\mathrm{BP2})}}{(\theta_L)_{i2}^{(\mathrm{BP1})}}   \sim \epsilon^{-3}\ , \quad
\frac{(\theta_L)_{i3}^{(\mathrm{BP2})}}{(\theta_L)_{i3}^{(\mathrm{BP1})}}   \sim \epsilon^{-5}\ ,
\end{align}
(see Eq.\,\eqref{eq:FNangle}). Consequently, the branching fractions into the $\overline{\nu}_\tau$ modes are enhanced by a factor of $\epsilon^{-4}$ in BP2 compared to BP1, dominating the $\overline{\nu}$ modes in BP2.

\begin{figure}[t!]
\centering
\subcaptionbox{BP1. 
\label{fig:BP1ratio}}
{\includegraphics[width=0.49\textwidth]{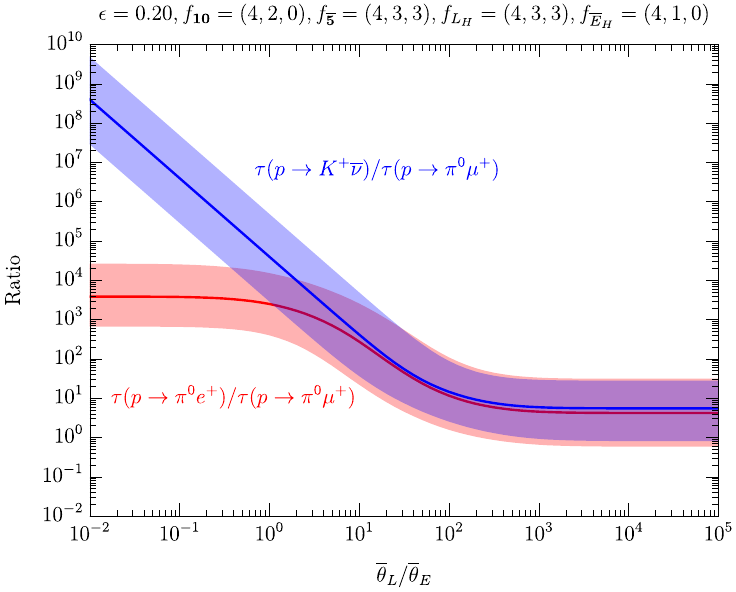}}
\subcaptionbox{BP2.
\label{fig:BP2ratio}}{\includegraphics[width=0.49\textwidth]{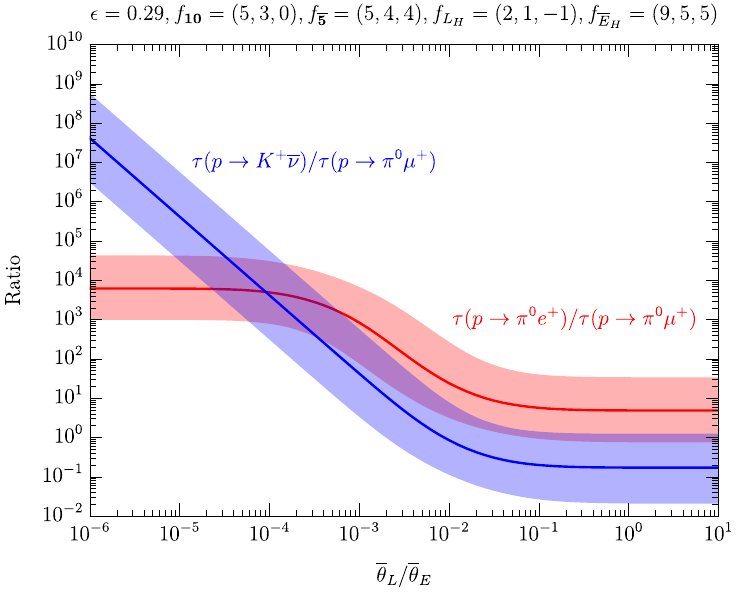}}
\caption{The lifetime ratios 
$\tau(p \to \pi^0 e^+)/\tau(p \to \pi^0 \mu^+)$
and 
$\tau(p \to K^+ \overline{\nu})/\tau(p \to \pi^0 \mu^+)$.
Figure (a) and (b) show the result for BP1 and BP2, 
respectively.
}
\label{fig:ratio}
\end{figure}

In Fig.\,\ref{fig:ratio}, 
we present the lifetime ratios 
$\tau(p \to \pi^0 e^+)/\tau(p \to \pi^0 \mu^+)$
and 
$\tau(p \to K^+ \overline{\nu})/\tau(p \to \pi^0 \mu^+)$
in the presence of both  $\mathcal{O}^{(1)}$ and $\mathcal{O}^{(2)}$ decay operators. 
These ratios are expressed as functions of the ratio between lepton doublet and singlet mixing angles, $\overline{\theta}_L/\overline{\theta}_E$ as described in Eq.\,\eqref{eq:thetabar}.
Our analysis reveals that the proton typically decays into $\pi^0 \mu^+$ rather than $\pi^0 e^+$ across the entire range of $\overline{\theta}_L/\overline{\theta}_E$ in both benchmark points. 
Conversely, 
the behavior of the 
$K^+ \overline{\nu}$ mode varies depending on the benchmark point. 
In BP1,
the proton is more likely to decay into
$p \to \pi^0 \mu^+$ 
than $K^+ \overline{\nu}$ for all values of 
$\overline{\theta}_L/\overline{\theta}_E$.
In BP2,
however, 
there exists a range of $\overline{\theta}_L/\overline{\theta}_E$ where the proton is more likely to decay into $K^+ \overline{\nu}$.
This behavior is influenced by $\overline{\theta}_L/\overline{\theta}_E$.
Other $\overline{\nu}$ decay modes associated with different mesons, such as $\pi$,  
have similar behaviors to 
that of
the $K^+ \overline{\nu}$ mode.

\section{Conclusions and Outlooks}

In this paper, 
we examined nucleon decay predictions within the fake GUT model based on $\mathrm{SU}(5) \times \mathrm{U}(2)_H$, incorporating the Froggatt-Nielsen (FN) mechanism. 
Our findings reveal that the flavor structure induced by the FN mechanism has a significant influence on nucleon decay modes. 
Due to the FN mechanism, 
each SU(5) multiplet includes substantial contributions from the second and third generations of SM leptons. 
Consequently, nucleon decays into $\mu^+$ and $\overline{\nu}_{\mu,\tau}$ become more probable. 
We identified the $p \to \pi^0 \mu^+$ mode as a prominent discovery channel. 
Moreover, 
the fake GUT model with the FN mechanism predicts a variety of nucleon decay channels, 
such as 
$p\to \pi^+ \overline{\nu}$, $n \to \pi^- \mu^+$, 
$n\to \pi^0 \overline{\nu}$,
and $p \to K^+ \overline{\nu}$. This contrasts with the conventional GUT framework, 
where the dominant mode is $p \to \pi^0 e^+$.

Our results demonstrate that
the FN charge assignments substantially affect the branching fractions of nucleon decay. 
We have illustrated these effects using two benchmark points, 
BP1 and BP2, 
which show clear distinctions in the predicted decay modes. 
Notably, 
several nucleon decay modes fall within the detection capabilities of upcoming experiments. 
Our findings also motivate experimental searches for less-explored decay modes. 
Furthermore, they indicate that nucleon decay searches can provide insights not only into baryon number violation but also into the origin of the flavor structure.

Future work will include a comprehensive survey of FN charge assignments and the impact of the nucleon decay\,\cite{ISW2024}. 
Additionally, 
we plan to extend this analysis to other models of flavor structures than the FN mechanism.
It is also interesting to apply the present analysis on the FN mechanism
to broader GUT models such as supersymmetric GUTs, 
where dimension-five operators may offer further insights into the impact of flavor symmetries on nucleon decay.

\section*{Acknowledgments}
This work is supported by Grant-in-Aid for Scientific Research from the Ministry of Education, Culture, Sports, Science, and Technology (MEXT), Japan, 20H01895 and 20H05860  (S.S.),
21H04471 and 22K03615 (M.I.) and by World Premier International Research Center Initiative (WPI), MEXT, Japan. 
This work is supported by JST SPRING Grant Number JPMJSP2108 and ANRI fellowship (K.W.).

\bibliographystyle{apsrev4-1}
\bibliography{bibtex}

\begin{thebibliography}{40}%
\makeatletter
\providecommand \@ifxundefined [1]{%
 \@ifx{#1\undefined}
}%
\providecommand \@ifnum [1]{%
 \ifnum #1\expandafter \@firstoftwo
 \else \expandafter \@secondoftwo
 \fi
}%
\providecommand \@ifx [1]{%
 \ifx #1\expandafter \@firstoftwo
 \else \expandafter \@secondoftwo
 \fi
}%
\providecommand \natexlab [1]{#1}%
\providecommand \enquote  [1]{``#1''}%
\providecommand \bibnamefont  [1]{#1}%
\providecommand \bibfnamefont [1]{#1}%
\providecommand \citenamefont [1]{#1}%
\providecommand \href@noop [0]{\@secondoftwo}%
\providecommand \href [0]{\begingroup \@sanitize@url \@href}%
\providecommand \@href[1]{\@@startlink{#1}\@@href}%
\providecommand \@@href[1]{\endgroup#1\@@endlink}%
\providecommand \@sanitize@url [0]{\catcode `\\12\catcode `\$12\catcode `\&12\catcode `\#12\catcode `\^12\catcode `\_12\catcode `\%12\relax}%
\providecommand \@@startlink[1]{}%
\providecommand \@@endlink[0]{}%
\providecommand \url  [0]{\begingroup\@sanitize@url \@url }%
\providecommand \@url [1]{\endgroup\@href {#1}{\urlprefix }}%
\providecommand \urlprefix  [0]{URL }%
\providecommand \Eprint [0]{\href }%
\providecommand \doibase [0]{http://dx.doi.org/}%
\providecommand \selectlanguage [0]{\@gobble}%
\providecommand \bibinfo  [0]{\@secondoftwo}%
\providecommand \bibfield  [0]{\@secondoftwo}%
\providecommand \translation [1]{[#1]}%
\providecommand \BibitemOpen [0]{}%
\providecommand \bibitemStop [0]{}%
\providecommand \bibitemNoStop [0]{.\EOS\space}%
\providecommand \EOS [0]{\spacefactor3000\relax}%
\providecommand \BibitemShut  [1]{\csname bibitem#1\endcsname}%
\let\auto@bib@innerbib\@empty
\bibitem [{\citenamefont {Georgi}\ and\ \citenamefont {Glashow}(1974)}]{Georgi:1974sy}%
  \BibitemOpen
  \bibfield  {author} {\bibinfo {author} {\bibfnamefont {H.}~\bibnamefont {Georgi}}\ and\ \bibinfo {author} {\bibfnamefont {S.~L.}\ \bibnamefont {Glashow}},\ }\href {\doibase 10.1103/PhysRevLett.32.438} {\bibfield  {journal} {\bibinfo  {journal} {Phys. Rev. Lett.}\ }\textbf {\bibinfo {volume} {32}},\ \bibinfo {pages} {438} (\bibinfo {year} {1974})}\BibitemShut {NoStop}%
\bibitem [{\citenamefont {Georgi}\ \emph {et~al.}(1974)\citenamefont {Georgi}, \citenamefont {Quinn},\ and\ \citenamefont {Weinberg}}]{Georgi:1974yf}%
  \BibitemOpen
  \bibfield  {author} {\bibinfo {author} {\bibfnamefont {H.}~\bibnamefont {Georgi}}, \bibinfo {author} {\bibfnamefont {H.~R.}\ \bibnamefont {Quinn}}, \ and\ \bibinfo {author} {\bibfnamefont {S.}~\bibnamefont {Weinberg}},\ }\href {\doibase 10.1103/PhysRevLett.33.451} {\bibfield  {journal} {\bibinfo  {journal} {Phys. Rev. Lett.}\ }\textbf {\bibinfo {volume} {33}},\ \bibinfo {pages} {451} (\bibinfo {year} {1974})}\BibitemShut {NoStop}%
\bibitem [{\citenamefont {Buras}\ \emph {et~al.}(1978)\citenamefont {Buras}, \citenamefont {Ellis}, \citenamefont {Gaillard},\ and\ \citenamefont {Nanopoulos}}]{Buras:1977yy}%
  \BibitemOpen
  \bibfield  {author} {\bibinfo {author} {\bibfnamefont {A.~J.}\ \bibnamefont {Buras}}, \bibinfo {author} {\bibfnamefont {J.~R.}\ \bibnamefont {Ellis}}, \bibinfo {author} {\bibfnamefont {M.~K.}\ \bibnamefont {Gaillard}}, \ and\ \bibinfo {author} {\bibfnamefont {D.~V.}\ \bibnamefont {Nanopoulos}},\ }\href {\doibase 10.1016/0550-3213(78)90214-6} {\bibfield  {journal} {\bibinfo  {journal} {Nucl. Phys. B}\ }\textbf {\bibinfo {volume} {135}},\ \bibinfo {pages} {66} (\bibinfo {year} {1978})}\BibitemShut {NoStop}%
\bibitem [{\citenamefont {Navas}\ \emph {et~al.}(2024)\citenamefont {Navas} \emph {et~al.}}]{ParticleDataGroup:2024cfk}%
  \BibitemOpen
  \bibfield  {author} {\bibinfo {author} {\bibfnamefont {S.}~\bibnamefont {Navas}} \emph {et~al.} (\bibinfo {collaboration} {Particle Data Group}),\ }\href {\doibase 10.1103/PhysRevD.110.030001} {\bibfield  {journal} {\bibinfo  {journal} {Phys. Rev. D}\ }\textbf {\bibinfo {volume} {110}},\ \bibinfo {pages} {030001} (\bibinfo {year} {2024})}\BibitemShut {NoStop}%
\bibitem [{\citenamefont {Takenaka}\ \emph {et~al.}(2020)\citenamefont {Takenaka} \emph {et~al.}}]{Super-Kamiokande:2020wjk}%
  \BibitemOpen
  \bibfield  {author} {\bibinfo {author} {\bibfnamefont {A.}~\bibnamefont {Takenaka}} \emph {et~al.} (\bibinfo {collaboration} {Super-Kamiokande}),\ }\href {\doibase 10.1103/PhysRevD.102.112011} {\bibfield  {journal} {\bibinfo  {journal} {Phys. Rev. D}\ }\textbf {\bibinfo {volume} {102}},\ \bibinfo {pages} {112011} (\bibinfo {year} {2020})},\ \Eprint {http://arxiv.org/abs/2010.16098} {arXiv:2010.16098 [hep-ex]} \BibitemShut {NoStop}%
\bibitem [{\citenamefont {Abe}\ \emph {et~al.}(2018)\citenamefont {Abe} \emph {et~al.}}]{Hyper-Kamiokande:2018ofw}%
  \BibitemOpen
  \bibfield  {author} {\bibinfo {author} {\bibfnamefont {K.}~\bibnamefont {Abe}} \emph {et~al.} (\bibinfo {collaboration} {Hyper-Kamiokande}),\ }\href@noop {} {\  (\bibinfo {year} {2018})},\ \Eprint {http://arxiv.org/abs/1805.04163} {arXiv:1805.04163 [physics.ins-det]} \BibitemShut {NoStop}%
\bibitem [{\citenamefont {An}\ \emph {et~al.}(2016)\citenamefont {An} \emph {et~al.}}]{JUNO:2015zny}%
  \BibitemOpen
  \bibfield  {author} {\bibinfo {author} {\bibfnamefont {F.}~\bibnamefont {An}} \emph {et~al.} (\bibinfo {collaboration} {JUNO}),\ }\href {\doibase 10.1088/0954-3899/43/3/030401} {\bibfield  {journal} {\bibinfo  {journal} {J. Phys. G}\ }\textbf {\bibinfo {volume} {43}},\ \bibinfo {pages} {030401} (\bibinfo {year} {2016})},\ \Eprint {http://arxiv.org/abs/1507.05613} {arXiv:1507.05613 [physics.ins-det]} \BibitemShut {NoStop}%
\bibitem [{\citenamefont {Abi}\ \emph {et~al.}(2021)\citenamefont {Abi} \emph {et~al.}}]{DUNE:2020fgq}%
  \BibitemOpen
  \bibfield  {author} {\bibinfo {author} {\bibfnamefont {B.}~\bibnamefont {Abi}} \emph {et~al.} (\bibinfo {collaboration} {DUNE}),\ }\href {\doibase 10.1140/epjc/s10052-021-09007-w} {\bibfield  {journal} {\bibinfo  {journal} {Eur. Phys. J. C}\ }\textbf {\bibinfo {volume} {81}},\ \bibinfo {pages} {322} (\bibinfo {year} {2021})},\ \Eprint {http://arxiv.org/abs/2008.12769} {arXiv:2008.12769 [hep-ex]} \BibitemShut {NoStop}%
\bibitem [{\citenamefont {Ibe}\ \emph {et~al.}(2019)\citenamefont {Ibe}, \citenamefont {Shirai}, \citenamefont {Suzuki},\ and\ \citenamefont {Yanagida}}]{Ibe:2019ifm}%
  \BibitemOpen
  \bibfield  {author} {\bibinfo {author} {\bibfnamefont {M.}~\bibnamefont {Ibe}}, \bibinfo {author} {\bibfnamefont {S.}~\bibnamefont {Shirai}}, \bibinfo {author} {\bibfnamefont {M.}~\bibnamefont {Suzuki}}, \ and\ \bibinfo {author} {\bibfnamefont {T.~T.}\ \bibnamefont {Yanagida}},\ }\href {\doibase 10.1103/PhysRevD.100.055024} {\bibfield  {journal} {\bibinfo  {journal} {Phys. Rev. D}\ }\textbf {\bibinfo {volume} {100}},\ \bibinfo {pages} {055024} (\bibinfo {year} {2019})},\ \Eprint {http://arxiv.org/abs/1906.02977} {arXiv:1906.02977 [hep-ph]} \BibitemShut {NoStop}%
\bibitem [{\citenamefont {Ibe}\ \emph {et~al.}(2022)\citenamefont {Ibe}, \citenamefont {Shirai}, \citenamefont {Suzuki}, \citenamefont {Watanabe},\ and\ \citenamefont {Yanagida}}]{Ibe:2022ock}%
  \BibitemOpen
  \bibfield  {author} {\bibinfo {author} {\bibfnamefont {M.}~\bibnamefont {Ibe}}, \bibinfo {author} {\bibfnamefont {S.}~\bibnamefont {Shirai}}, \bibinfo {author} {\bibfnamefont {M.}~\bibnamefont {Suzuki}}, \bibinfo {author} {\bibfnamefont {K.}~\bibnamefont {Watanabe}}, \ and\ \bibinfo {author} {\bibfnamefont {T.~T.}\ \bibnamefont {Yanagida}},\ }\href {\doibase 10.1007/JHEP07(2022)087} {\bibfield  {journal} {\bibinfo  {journal} {JHEP}\ }\textbf {\bibinfo {volume} {07}},\ \bibinfo {pages} {087} (\bibinfo {year} {2022})},\ \Eprint {http://arxiv.org/abs/2205.01336} {arXiv:2205.01336 [hep-ph]} \BibitemShut {NoStop}%
\bibitem [{\citenamefont {Hall}\ and\ \citenamefont {Harigaya}(2018)}]{Hall:2018let}%
  \BibitemOpen
  \bibfield  {author} {\bibinfo {author} {\bibfnamefont {L.~J.}\ \bibnamefont {Hall}}\ and\ \bibinfo {author} {\bibfnamefont {K.}~\bibnamefont {Harigaya}},\ }\href {\doibase 10.1007/JHEP10(2018)130} {\bibfield  {journal} {\bibinfo  {journal} {JHEP}\ }\textbf {\bibinfo {volume} {10}},\ \bibinfo {pages} {130} (\bibinfo {year} {2018})},\ \Eprint {http://arxiv.org/abs/1803.08119} {arXiv:1803.08119 [hep-ph]} \BibitemShut {NoStop}%
\bibitem [{\citenamefont {Hall}\ and\ \citenamefont {Harigaya}(2019)}]{Hall:2019qwx}%
  \BibitemOpen
  \bibfield  {author} {\bibinfo {author} {\bibfnamefont {L.~J.}\ \bibnamefont {Hall}}\ and\ \bibinfo {author} {\bibfnamefont {K.}~\bibnamefont {Harigaya}},\ }\href {\doibase 10.1007/JHEP11(2019)033} {\bibfield  {journal} {\bibinfo  {journal} {JHEP}\ }\textbf {\bibinfo {volume} {11}},\ \bibinfo {pages} {033} (\bibinfo {year} {2019})},\ \Eprint {http://arxiv.org/abs/1905.12722} {arXiv:1905.12722 [hep-ph]} \BibitemShut {NoStop}%
\bibitem [{\citenamefont {Fritzsch}\ and\ \citenamefont {Xing}(2000)}]{Fritzsch:1999ee}%
  \BibitemOpen
  \bibfield  {author} {\bibinfo {author} {\bibfnamefont {H.}~\bibnamefont {Fritzsch}}\ and\ \bibinfo {author} {\bibfnamefont {Z.-z.}\ \bibnamefont {Xing}},\ }\href {\doibase 10.1016/S0146-6410(00)00102-2} {\bibfield  {journal} {\bibinfo  {journal} {Prog. Part. Nucl. Phys.}\ }\textbf {\bibinfo {volume} {45}},\ \bibinfo {pages} {1} (\bibinfo {year} {2000})},\ \Eprint {http://arxiv.org/abs/hep-ph/9912358} {arXiv:hep-ph/9912358} \BibitemShut {NoStop}%
\bibitem [{\citenamefont {Altarelli}\ and\ \citenamefont {Feruglio}(2010)}]{Altarelli:2010gt}%
  \BibitemOpen
  \bibfield  {author} {\bibinfo {author} {\bibfnamefont {G.}~\bibnamefont {Altarelli}}\ and\ \bibinfo {author} {\bibfnamefont {F.}~\bibnamefont {Feruglio}},\ }\href {\doibase 10.1103/RevModPhys.82.2701} {\bibfield  {journal} {\bibinfo  {journal} {Rev. Mod. Phys.}\ }\textbf {\bibinfo {volume} {82}},\ \bibinfo {pages} {2701} (\bibinfo {year} {2010})},\ \Eprint {http://arxiv.org/abs/1002.0211} {arXiv:1002.0211 [hep-ph]} \BibitemShut {NoStop}%
\bibitem [{\citenamefont {Froggatt}\ and\ \citenamefont {Nielsen}(1979)}]{Froggatt:1978nt}%
  \BibitemOpen
  \bibfield  {author} {\bibinfo {author} {\bibfnamefont {C.~D.}\ \bibnamefont {Froggatt}}\ and\ \bibinfo {author} {\bibfnamefont {H.~B.}\ \bibnamefont {Nielsen}},\ }\href {\doibase 10.1016/0550-3213(79)90316-X} {\bibfield  {journal} {\bibinfo  {journal} {Nucl. Phys. B}\ }\textbf {\bibinfo {volume} {147}},\ \bibinfo {pages} {277} (\bibinfo {year} {1979})}\BibitemShut {NoStop}%
\bibitem [{\citenamefont {'t~Hooft}\ \emph {et~al.}(1980)\citenamefont {'t~Hooft}, \citenamefont {Itzykson}, \citenamefont {Jaffe}, \citenamefont {Lehmann}, \citenamefont {Mitter}, \citenamefont {Singer},\ and\ \citenamefont {Stora}}]{tHooft:1980xss}%
  \BibitemOpen
  \bibinfo {editor} {\bibfnamefont {G.}~\bibnamefont {'t~Hooft}}, \bibinfo {editor} {\bibfnamefont {C.}~\bibnamefont {Itzykson}}, \bibinfo {editor} {\bibfnamefont {A.}~\bibnamefont {Jaffe}}, \bibinfo {editor} {\bibfnamefont {H.}~\bibnamefont {Lehmann}}, \bibinfo {editor} {\bibfnamefont {P.~K.}\ \bibnamefont {Mitter}}, \bibinfo {editor} {\bibfnamefont {I.~M.}\ \bibnamefont {Singer}}, \ and\ \bibinfo {editor} {\bibfnamefont {R.}~\bibnamefont {Stora}},\ eds.,\ \href {\doibase 10.1007/978-1-4684-7571-5} {\emph {\bibinfo {title} {{Recent Developments in Gauge Theories. Proceedings, Nato Advanced Study Institute, Cargese, France, August 26 - September 8, 1979}}}},\ Vol.~\bibinfo {volume} {59}\ (\bibinfo {year} {1980})\BibitemShut {NoStop}%
\bibitem [{\citenamefont {Minkowski}(1977)}]{Minkowski:1977sc}%
  \BibitemOpen
  \bibfield  {author} {\bibinfo {author} {\bibfnamefont {P.}~\bibnamefont {Minkowski}},\ }\href {\doibase 10.1016/0370-2693(77)90435-X} {\bibfield  {journal} {\bibinfo  {journal} {Phys. Lett. B}\ }\textbf {\bibinfo {volume} {67}},\ \bibinfo {pages} {421} (\bibinfo {year} {1977})}\BibitemShut {NoStop}%
\bibitem [{\citenamefont {Yanagida}(1979{\natexlab{a}})}]{Yanagida:1979as}%
  \BibitemOpen
  \bibfield  {author} {\bibinfo {author} {\bibfnamefont {T.}~\bibnamefont {Yanagida}},\ }\href@noop {} {\bibfield  {journal} {\bibinfo  {journal} {Conf. Proc. C}\ }\textbf {\bibinfo {volume} {7902131}},\ \bibinfo {pages} {95} (\bibinfo {year} {1979}{\natexlab{a}})}\BibitemShut {NoStop}%
\bibitem [{\citenamefont {Yanagida}(1979{\natexlab{b}})}]{Yanagida:1979gs}%
  \BibitemOpen
  \bibfield  {author} {\bibinfo {author} {\bibfnamefont {T.}~\bibnamefont {Yanagida}},\ }\href {\doibase 10.1103/PhysRevD.20.2986} {\bibfield  {journal} {\bibinfo  {journal} {Phys. Rev. D}\ }\textbf {\bibinfo {volume} {20}},\ \bibinfo {pages} {2986} (\bibinfo {year} {1979}{\natexlab{b}})}\BibitemShut {NoStop}%
\bibitem [{\citenamefont {Gell-Mann}\ \emph {et~al.}(1979)\citenamefont {Gell-Mann}, \citenamefont {Ramond},\ and\ \citenamefont {Slansky}}]{Gell-Mann:1979vob}%
  \BibitemOpen
  \bibfield  {author} {\bibinfo {author} {\bibfnamefont {M.}~\bibnamefont {Gell-Mann}}, \bibinfo {author} {\bibfnamefont {P.}~\bibnamefont {Ramond}}, \ and\ \bibinfo {author} {\bibfnamefont {R.}~\bibnamefont {Slansky}},\ }\href@noop {} {\bibfield  {journal} {\bibinfo  {journal} {Conf. Proc. C}\ }\textbf {\bibinfo {volume} {790927}},\ \bibinfo {pages} {315} (\bibinfo {year} {1979})},\ \Eprint {http://arxiv.org/abs/1306.4669} {arXiv:1306.4669 [hep-th]} \BibitemShut {NoStop}%
\bibitem [{\citenamefont {Glashow}(1980)}]{Glashow:1979nm}%
  \BibitemOpen
  \bibfield  {author} {\bibinfo {author} {\bibfnamefont {S.~L.}\ \bibnamefont {Glashow}},\ }\href {\doibase 10.1007/978-1-4684-7197-7_15} {\bibfield  {journal} {\bibinfo  {journal} {NATO Sci. Ser. B}\ }\textbf {\bibinfo {volume} {61}},\ \bibinfo {pages} {687} (\bibinfo {year} {1980})}\BibitemShut {NoStop}%
\bibitem [{\citenamefont {Mohapatra}\ and\ \citenamefont {Senjanovic}(1980)}]{Mohapatra:1979ia}%
  \BibitemOpen
  \bibfield  {author} {\bibinfo {author} {\bibfnamefont {R.~N.}\ \bibnamefont {Mohapatra}}\ and\ \bibinfo {author} {\bibfnamefont {G.}~\bibnamefont {Senjanovic}},\ }\href {\doibase 10.1103/PhysRevLett.44.912} {\bibfield  {journal} {\bibinfo  {journal} {Phys. Rev. Lett.}\ }\textbf {\bibinfo {volume} {44}},\ \bibinfo {pages} {912} (\bibinfo {year} {1980})}\BibitemShut {NoStop}%
\bibitem [{\citenamefont {Sato}\ and\ \citenamefont {Yanagida}(1998)}]{Sato:1997hv}%
  \BibitemOpen
  \bibfield  {author} {\bibinfo {author} {\bibfnamefont {J.}~\bibnamefont {Sato}}\ and\ \bibinfo {author} {\bibfnamefont {T.}~\bibnamefont {Yanagida}},\ }\href {\doibase 10.1016/S0370-2693(98)00510-3} {\bibfield  {journal} {\bibinfo  {journal} {Phys. Lett. B}\ }\textbf {\bibinfo {volume} {430}},\ \bibinfo {pages} {127} (\bibinfo {year} {1998})},\ \Eprint {http://arxiv.org/abs/hep-ph/9710516} {arXiv:hep-ph/9710516} \BibitemShut {NoStop}%
\bibitem [{\citenamefont {Babu}\ \emph {et~al.}(2017)\citenamefont {Babu}, \citenamefont {Khanov},\ and\ \citenamefont {Saad}}]{Babu:2016aro}%
  \BibitemOpen
  \bibfield  {author} {\bibinfo {author} {\bibfnamefont {K.~S.}\ \bibnamefont {Babu}}, \bibinfo {author} {\bibfnamefont {A.}~\bibnamefont {Khanov}}, \ and\ \bibinfo {author} {\bibfnamefont {S.}~\bibnamefont {Saad}},\ }\href {\doibase 10.1103/PhysRevD.95.055014} {\bibfield  {journal} {\bibinfo  {journal} {Phys. Rev. D}\ }\textbf {\bibinfo {volume} {95}},\ \bibinfo {pages} {055014} (\bibinfo {year} {2017})},\ \Eprint {http://arxiv.org/abs/1612.07787} {arXiv:1612.07787 [hep-ph]} \BibitemShut {NoStop}%
\bibitem [{\citenamefont {Workman}\ \emph {et~al.}(2022)\citenamefont {Workman} \emph {et~al.}}]{ParticleDataGroup:2022pth}%
  \BibitemOpen
  \bibfield  {author} {\bibinfo {author} {\bibfnamefont {R.~L.}\ \bibnamefont {Workman}} \emph {et~al.} (\bibinfo {collaboration} {Particle Data Group}),\ }\href {\doibase 10.1093/ptep/ptac097} {\bibfield  {journal} {\bibinfo  {journal} {PTEP}\ }\textbf {\bibinfo {volume} {2022}},\ \bibinfo {pages} {083C01} (\bibinfo {year} {2022})}\BibitemShut {NoStop}%
\bibitem [{\citenamefont {Buttazzo}\ \emph {et~al.}(2013)\citenamefont {Buttazzo}, \citenamefont {Degrassi}, \citenamefont {Giardino}, \citenamefont {Giudice}, \citenamefont {Sala}, \citenamefont {Salvio},\ and\ \citenamefont {Strumia}}]{Buttazzo:2013uya}%
  \BibitemOpen
  \bibfield  {author} {\bibinfo {author} {\bibfnamefont {D.}~\bibnamefont {Buttazzo}}, \bibinfo {author} {\bibfnamefont {G.}~\bibnamefont {Degrassi}}, \bibinfo {author} {\bibfnamefont {P.~P.}\ \bibnamefont {Giardino}}, \bibinfo {author} {\bibfnamefont {G.~F.}\ \bibnamefont {Giudice}}, \bibinfo {author} {\bibfnamefont {F.}~\bibnamefont {Sala}}, \bibinfo {author} {\bibfnamefont {A.}~\bibnamefont {Salvio}}, \ and\ \bibinfo {author} {\bibfnamefont {A.}~\bibnamefont {Strumia}},\ }\href {\doibase 10.1007/JHEP12(2013)089} {\bibfield  {journal} {\bibinfo  {journal} {JHEP}\ }\textbf {\bibinfo {volume} {12}},\ \bibinfo {pages} {089} (\bibinfo {year} {2013})},\ \Eprint {http://arxiv.org/abs/1307.3536} {arXiv:1307.3536 [hep-ph]} \BibitemShut {NoStop}%
\bibitem [{\citenamefont {Chetyrkin}\ \emph {et~al.}(1997)\citenamefont {Chetyrkin}, \citenamefont {Kniehl},\ and\ \citenamefont {Steinhauser}}]{Chetyrkin:1997sg}%
  \BibitemOpen
  \bibfield  {author} {\bibinfo {author} {\bibfnamefont {K.~G.}\ \bibnamefont {Chetyrkin}}, \bibinfo {author} {\bibfnamefont {B.~A.}\ \bibnamefont {Kniehl}}, \ and\ \bibinfo {author} {\bibfnamefont {M.}~\bibnamefont {Steinhauser}},\ }\href {\doibase 10.1103/PhysRevLett.79.2184} {\bibfield  {journal} {\bibinfo  {journal} {Phys. Rev. Lett.}\ }\textbf {\bibinfo {volume} {79}},\ \bibinfo {pages} {2184} (\bibinfo {year} {1997})},\ \Eprint {http://arxiv.org/abs/hep-ph/9706430} {arXiv:hep-ph/9706430} \BibitemShut {NoStop}%
\bibitem [{\citenamefont {Nagata}\ and\ \citenamefont {Shirai}(2014)}]{Nagata:2013sba}%
  \BibitemOpen
  \bibfield  {author} {\bibinfo {author} {\bibfnamefont {N.}~\bibnamefont {Nagata}}\ and\ \bibinfo {author} {\bibfnamefont {S.}~\bibnamefont {Shirai}},\ }\href {\doibase 10.1007/JHEP03(2014)049} {\bibfield  {journal} {\bibinfo  {journal} {JHEP}\ }\textbf {\bibinfo {volume} {03}},\ \bibinfo {pages} {049} (\bibinfo {year} {2014})},\ \Eprint {http://arxiv.org/abs/1312.7854} {arXiv:1312.7854 [hep-ph]} \BibitemShut {NoStop}%
\bibitem [{\citenamefont {Abbott}\ and\ \citenamefont {Wise}(1980)}]{Abbott:1980zj}%
  \BibitemOpen
  \bibfield  {author} {\bibinfo {author} {\bibfnamefont {L.~F.}\ \bibnamefont {Abbott}}\ and\ \bibinfo {author} {\bibfnamefont {M.~B.}\ \bibnamefont {Wise}},\ }\href {\doibase 10.1103/PhysRevD.22.2208} {\bibfield  {journal} {\bibinfo  {journal} {Phys. Rev. D}\ }\textbf {\bibinfo {volume} {22}},\ \bibinfo {pages} {2208} (\bibinfo {year} {1980})}\BibitemShut {NoStop}%
\bibitem [{\citenamefont {Nihei}\ and\ \citenamefont {Arafune}(1995)}]{Nihei:1994tx}%
  \BibitemOpen
  \bibfield  {author} {\bibinfo {author} {\bibfnamefont {T.}~\bibnamefont {Nihei}}\ and\ \bibinfo {author} {\bibfnamefont {J.}~\bibnamefont {Arafune}},\ }\href {\doibase 10.1143/PTP.93.665} {\bibfield  {journal} {\bibinfo  {journal} {Prog. Theor. Phys.}\ }\textbf {\bibinfo {volume} {93}},\ \bibinfo {pages} {665} (\bibinfo {year} {1995})},\ \Eprint {http://arxiv.org/abs/hep-ph/9412325} {arXiv:hep-ph/9412325} \BibitemShut {NoStop}%
\bibitem [{\citenamefont {Yoo}\ \emph {et~al.}(2022)\citenamefont {Yoo}, \citenamefont {Aoki}, \citenamefont {Boyle}, \citenamefont {Izubuchi}, \citenamefont {Soni},\ and\ \citenamefont {Syritsyn}}]{Yoo:2021gql}%
  \BibitemOpen
  \bibfield  {author} {\bibinfo {author} {\bibfnamefont {J.-S.}\ \bibnamefont {Yoo}}, \bibinfo {author} {\bibfnamefont {Y.}~\bibnamefont {Aoki}}, \bibinfo {author} {\bibfnamefont {P.}~\bibnamefont {Boyle}}, \bibinfo {author} {\bibfnamefont {T.}~\bibnamefont {Izubuchi}}, \bibinfo {author} {\bibfnamefont {A.}~\bibnamefont {Soni}}, \ and\ \bibinfo {author} {\bibfnamefont {S.}~\bibnamefont {Syritsyn}},\ }\href {\doibase 10.1103/PhysRevD.105.074501} {\bibfield  {journal} {\bibinfo  {journal} {Phys. Rev. D}\ }\textbf {\bibinfo {volume} {105}},\ \bibinfo {pages} {074501} (\bibinfo {year} {2022})},\ \Eprint {http://arxiv.org/abs/2111.01608} {arXiv:2111.01608 [hep-lat]} \BibitemShut {NoStop}%
\bibitem [{\citenamefont {Aoki}\ \emph {et~al.}(2017)\citenamefont {Aoki}, \citenamefont {Izubuchi}, \citenamefont {Shintani},\ and\ \citenamefont {Soni}}]{Aoki:2017puj}%
  \BibitemOpen
  \bibfield  {author} {\bibinfo {author} {\bibfnamefont {Y.}~\bibnamefont {Aoki}}, \bibinfo {author} {\bibfnamefont {T.}~\bibnamefont {Izubuchi}}, \bibinfo {author} {\bibfnamefont {E.}~\bibnamefont {Shintani}}, \ and\ \bibinfo {author} {\bibfnamefont {A.}~\bibnamefont {Soni}},\ }\href {\doibase 10.1103/PhysRevD.96.014506} {\bibfield  {journal} {\bibinfo  {journal} {Phys. Rev. D}\ }\textbf {\bibinfo {volume} {96}},\ \bibinfo {pages} {014506} (\bibinfo {year} {2017})},\ \Eprint {http://arxiv.org/abs/1705.01338} {arXiv:1705.01338 [hep-lat]} \BibitemShut {NoStop}%
\bibitem [{\citenamefont {Machacek}(1979)}]{Machacek:1979tx}%
  \BibitemOpen
  \bibfield  {author} {\bibinfo {author} {\bibfnamefont {M.}~\bibnamefont {Machacek}},\ }\href {\doibase 10.1016/0550-3213(79)90325-0} {\bibfield  {journal} {\bibinfo  {journal} {Nucl. Phys. B}\ }\textbf {\bibinfo {volume} {159}},\ \bibinfo {pages} {37} (\bibinfo {year} {1979})}\BibitemShut {NoStop}%
\bibitem [{\citenamefont {Weinberg}(1979)}]{Weinberg:1979sa}%
  \BibitemOpen
  \bibfield  {author} {\bibinfo {author} {\bibfnamefont {S.}~\bibnamefont {Weinberg}},\ }\href {\doibase 10.1103/PhysRevLett.43.1566} {\bibfield  {journal} {\bibinfo  {journal} {Phys. Rev. Lett.}\ }\textbf {\bibinfo {volume} {43}},\ \bibinfo {pages} {1566} (\bibinfo {year} {1979})}\BibitemShut {NoStop}%
\bibitem [{\citenamefont {Abe}\ \emph {et~al.}(2014{\natexlab{a}})\citenamefont {Abe} \emph {et~al.}}]{Super-Kamiokande:2013rwg}%
  \BibitemOpen
  \bibfield  {author} {\bibinfo {author} {\bibfnamefont {K.}~\bibnamefont {Abe}} \emph {et~al.} (\bibinfo {collaboration} {Super-Kamiokande}),\ }\href {\doibase 10.1103/PhysRevLett.113.121802} {\bibfield  {journal} {\bibinfo  {journal} {Phys. Rev. Lett.}\ }\textbf {\bibinfo {volume} {113}},\ \bibinfo {pages} {121802} (\bibinfo {year} {2014}{\natexlab{a}})},\ \Eprint {http://arxiv.org/abs/1305.4391} {arXiv:1305.4391 [hep-ex]} \BibitemShut {NoStop}%
\bibitem [{\citenamefont {Kobayashi}\ \emph {et~al.}(2005)\citenamefont {Kobayashi} \emph {et~al.}}]{Super-Kamiokande:2005lev}%
  \BibitemOpen
  \bibfield  {author} {\bibinfo {author} {\bibfnamefont {K.}~\bibnamefont {Kobayashi}} \emph {et~al.} (\bibinfo {collaboration} {Super-Kamiokande}),\ }\href {\doibase 10.1103/PhysRevD.72.052007} {\bibfield  {journal} {\bibinfo  {journal} {Phys. Rev. D}\ }\textbf {\bibinfo {volume} {72}},\ \bibinfo {pages} {052007} (\bibinfo {year} {2005})},\ \Eprint {http://arxiv.org/abs/hep-ex/0502026} {arXiv:hep-ex/0502026} \BibitemShut {NoStop}%
\bibitem [{\citenamefont {Matsumoto}\ \emph {et~al.}(2022)\citenamefont {Matsumoto} \emph {et~al.}}]{Super-Kamiokande:2022egr}%
  \BibitemOpen
  \bibfield  {author} {\bibinfo {author} {\bibfnamefont {R.}~\bibnamefont {Matsumoto}} \emph {et~al.} (\bibinfo {collaboration} {Super-Kamiokande}),\ }\href {\doibase 10.1103/PhysRevD.106.072003} {\bibfield  {journal} {\bibinfo  {journal} {Phys. Rev. D}\ }\textbf {\bibinfo {volume} {106}},\ \bibinfo {pages} {072003} (\bibinfo {year} {2022})},\ \Eprint {http://arxiv.org/abs/2208.13188} {arXiv:2208.13188 [hep-ex]} \BibitemShut {NoStop}%
\bibitem [{\citenamefont {Abe}\ \emph {et~al.}(2014{\natexlab{b}})\citenamefont {Abe} \emph {et~al.}}]{Super-Kamiokande:2014otb}%
  \BibitemOpen
  \bibfield  {author} {\bibinfo {author} {\bibfnamefont {K.}~\bibnamefont {Abe}} \emph {et~al.} (\bibinfo {collaboration} {Super-Kamiokande}),\ }\href {\doibase 10.1103/PhysRevD.90.072005} {\bibfield  {journal} {\bibinfo  {journal} {Phys. Rev. D}\ }\textbf {\bibinfo {volume} {90}},\ \bibinfo {pages} {072005} (\bibinfo {year} {2014}{\natexlab{b}})},\ \Eprint {http://arxiv.org/abs/1408.1195} {arXiv:1408.1195 [hep-ex]} \BibitemShut {NoStop}%
\bibitem [{\citenamefont {Taniuchi}\ \emph {et~al.}(2024)\citenamefont {Taniuchi} \emph {et~al.}}]{Super-Kamiokande:2024qbv}%
  \BibitemOpen
  \bibfield  {author} {\bibinfo {author} {\bibfnamefont {N.}~\bibnamefont {Taniuchi}} \emph {et~al.} (\bibinfo {collaboration} {Super-Kamiokande}),\ }\href@noop {} {\  (\bibinfo {year} {2024})},\ \Eprint {http://arxiv.org/abs/2409.19633} {arXiv:2409.19633 [hep-ex]} \BibitemShut {NoStop}%
\bibitem [{\citenamefont {Ibe}\ \emph {et~al.}()\citenamefont {Ibe}, \citenamefont {Shirai},\ and\ \citenamefont {Watanabe}}]{ISW2024}%
  \BibitemOpen
  \bibfield  {author} {\bibinfo {author} {\bibfnamefont {M.}~\bibnamefont {Ibe}}, \bibinfo {author} {\bibfnamefont {S.}~\bibnamefont {Shirai}}, \ and\ \bibinfo {author} {\bibfnamefont {K.}~\bibnamefont {Watanabe}},\ }\href@noop {} {}\bibinfo {note} {$\phantom{X}$\!\!\!\!\!to appear}\BibitemShut {NoStop}%
\end{thebibliography}%

\end{document}